**Evidence for Emergent Kagome Spin Configuration with Concomitant Transverse and Longitudinal Spin-Glass Freezing in the Chemically Ordered M-type Hexaferrite BaFe$_{12}$O$_{19}$**

Keshav Kumar,[1] Anatoliy Senyshyn,[2] and Dhananjai Pandey[1,*]

[1]School of Materials Science and Technology, Indian Institute of Technology (Banaras Hindu University), Varanasi-221005, India.
[2]Forschungsneutronenquelle Heinze Maier-Leibnitz (FRM-II), Technische Universitat Munchen, Liechtenbergestrass 1, D-85747 Garching b. Munchen, Germany.

**Abstract**

Frustration effects in magnetic systems have traditionally been investigated considering pre-existing site-disorder or lattice geometry of the high-temperature paramagnetic phase. We present here evidence for emergence of geometrical frustration as a function of temperature due to spin canting in the long-range ordered (LRO) ferrimagnetic (FMI) phase of BaFe$_{12}$O$_{19}$ (BFO), an M-type hexaferrite of enormous technological applications. Results of neutron scattering and magnetic susceptibility studies on BFO are presented to show for the first time the emergence of highly degenerate kagome spin configuration for the basal plane spin component of BFO with concomitant freezing of transverse and longitudinal components of the spins leading to two spin-glass transitions in coexistence with the LRO FMI phase. Our results mimic the theoretical predictions for concentrated Heisenberg systems even though the source of frustration in BFO is the geometry of the lattice and not site-disorder. We believe that our findings will stimulate theoretical studies to unravel the physics of spin-glass transitions in LRO systems due to emergent geometrical frustration, an aspect that has remained unexplored so far. We also believe that this work will encourage further experimental studies in search of low temperature spin-glass transition(s) in LRO phases of various hexaferrites and even other LRO magnetic compounds with spins arranged on triangular, pyrochlore and spinel lattices without any substitutional disorder.

**1. Introduction**

The study of phase transitions in frustrated magnetic systems has been a frontline area of research in condensed matter physics and material sciences for decades [1–12]. The most commonly investigated source of frustration is randomly frozen-in substitutional disorder leading to a competition between ferromagnetic (FM) and antiferromagnetic (AFM) interactions [13–15]. Such a competition and randomness in certain situations can prevent the emergence of long-range ordered (LRO) phases and instead give rise to a unique spin-glass state in which the spins are randomly frozen below a frequency ($\omega$) and field (H) dependent peak, usually called as spin-glass freezing temperature ($T_f$), in the temperature dependence of magnetic susceptibility as a result of ergodic symmetry breaking [13–15]. Frustration may also arise due to nearest neighbour AFM interactions alone due to the geometry of the lattice [2,5,6,8,16] for spins arranged on the edge shared triangular (e.g. $YbZnGaO_4$) [10], corner shared triangular or kagome (e.g. $SrSn_2Fe_4O_{11}$, $Cd_2Cu_3(OH)_6(SO_4)24H_2O$, $BaNi_3(OH)_2(VO_4)_2$, $Pr_3Ga_5SiO_{14}$, $Ba_2Sn_2Ga_3ZnCr_7O_{22}$, $SrGa_{12-x}Cr_xO_{19}$) [17–22], pyrochlore (e.g. $Y_2Mo_2O_7$, $Lu_2Mo_2O_7$, $Tb_2Mo_2O_7$, $A_2Sb_2O_7$ with A = Mn, Co, Ni) [23–28], and spinel (e.g. $MAl_2O_4$ with M = Co, Fe, & Mn) [9,29] lattices. Such geometrically frustrated systems have evinced enormous attention in recent times as they exhibit exotic spin liquid (quantum as well as classical) [8,30], spin ice [1,8,31,32] and spin-glass transitions [17,27,33–37] even in the absence of any apparent substitutional disorder. Theoretical calculations seem to suggest that in the absence of any site-disorder, and hence randomness, the ground state of the geometrically frustrated systems has macroscopic degeneracy with no phase transition down to the absolute zero temperature [38–40]. However, it has also been shown that even an infinitesimal random disorder in the few body exchange interactions [40,41], caused by anisotropic exchange interactions due to nearest neighbour bond length variations [25,27,42,43] and/or magnetoelastic strains [41] or dipole-dipole interactions between uncompensated spin or spin clusters with intra-cluster geometrical frustration [42–46], can lift the degeneracy of the ground state and induce phase transitions to spin liquid [10,28], spin-glass [10,27,37] and other complex ordered phases [47,48].

Frustration resulting from the site-disorder or the geometry of the lattice considered so far in the literature pre-exisists in the paramagnetic high temperature phase. We present here results of magnetic susceptibility and neutron scattering studies on the LRO ferrimagnetic phase of $BaFe_{12}O_{19}$ (BFO), an M-type hexaferrite with enormous technological applications [49–51], which reveal emergence of kagome spin configurations below the room temperature for the basal plane spin component, as a result of spin canting, and a concomitant freezing of transverse and some part of the longitudinal component of the spins leading to two spin-glass transitions. This is the first evidence for a succession of two spin-glass transitions in coexistence with the LRO phase in a chemically ordered compound due to emergent geometrical frustration. Interestingly, our results mimic the theoretical predictions for conventional disordered Heisenberg systems near the percolation threshold limit [52] even though the source of frustration is geometry of the lattice and not site-disorder in our case. We believe that our findings will stimulate further theoretical and experimental studies in search of emergent geometrical frustration and its consequences in other hexaferrites [49] as well as other ordered magnetic materials with spins arranged on triangular, tetrahedral and kagome lattices [2,5–7].

## 2. Experimental Details and Analysis

Barium hexaferrite (BaFe$_{12}$O$_{19}$) was synthesized by the solid-state thermo-chemical reaction route details of which are given elsewhere [53]. Single crystals of BaFe$_{12}$O$_{19}$ were grown using conventional high temperature flux method using Na$_2$CO$_3$ as self-flux [54]. The as-grown crystals are of hexagonal platelet shape with 3-6 mm lateral dimension and ~0.5 to ~1.0 mm thickness. Both the powder and single-crystal samples were well characterized using synchrotron x-ray powder diffraction (SXRD) and laboratory based x-ray source data for their phase purity and crystal orientation, respectively [53,55]. The dc magnetization (M(T)) and ac susceptibility ($\chi(\omega, T)$) measurements in the 2 to 300K temperature (T) range were carried out in the warming cycle on zero-field cooled (ZFC) samples with dc and ac fields applied parallel (H//$c$) and perpendicular (H$\perp c$) to the c-axis of the single-crystal samples using a SQUID-based magnetometer (MPMS-3, Quantum Design, USA). The magnetic measurements on powder samples were carried out on a Physical Property Measurements System (PPMS, Dynacool, Quantum Design, USA). The temperature dependent neutron diffraction patterns of the powder samples were recorded at the high-resolution powder diffractometer SPODI at FRM II Hamburg, Germany [56].

The average magnetic and nuclear structures were refined by Rietveld method using the FullProf suite [57]. In the refinements, the background was modeled using linear interpolation between the points. The peak shape was modeled using Thompson-Cox-Hasting pseudo-Voigt function. The Full-Width at Half-Maxima (FWHM) of the peaks were modeled using the Caglioti equation [58,59]:

$$(FWHM)^2 = U\tan^2\theta + V\tan\theta + W \qquad \ldots\ldots(1)$$

During the refinement, scale factor, zero displacement, lattice parameters (a, b, c), positional coordinates (x, y, z) and thermal parameters (B) were allowed to vary while the occupancy of each ion was fixed at their nominal composition value. The initial input structural and thermal parameters were taken from ref. [60].

## 3. Results and Discussion
### A. Magnetic Anisotropy

Magnetic structure of M-type hexaferrites is very verge complex with as many as 12 $Fe^{3+}$ ions having three different coordinations, namely tetrahedral (TH), octahedral (OH) and trigonal bipyramid (TBP), of oxygen ions occupying five different crystallographic Wyckoff sites 2a (OH), 2b (TBP), $4f_1$ (TH), $4f_2$ (OH), and 12k (OH) [61]. Gorter [61] demonstrated an excellent agreement between the saturation magnetization calculated using a collinear magnetic structure and the experimental value of $40\mu_B$ per unit cell [61]. As per this model, all the $3dFe^{3+}$ spins in the M-type hexaferrites are Ising type with the spins at the 2a, 2b and 12k Wyckoff sites are all up while the spins at the $4f_1$ and $4f_2$ sites are all down along the c-axis. This spin configuration gives rise to the overall collinear ferrimagnetic structure with a net magnetic moment of $40\mu_B$ per unit cell which is close to what is observed experimentally [61].

The first signature of the deviation from the Gorter model of collinear Ising spins of BFO at higher temperatures is revealed through the magnetic anisotropy of the dc magnetization M(T). The variation of $M_{\perp c}(T)$ and $M_{//c}(T)$, measured during the warming cycle on a zero-field cooled (ZFC) single-crystal, with a magnetic field of 500 Oe applied perpendicular ($\perp$) and parallel (//) to the c-axis, respectively, is shown in Fig. 1(a). These plots were obtained by merging the results of measurements in the 300 K to 950 K and 2 K to 300 K temperature ranges using high and low-temperature attachments, respectively, in the PPMS system. To locate the LRO ferrimagnetic transition temperature, i.e., the Curie temperature ($T_c$), we use the first derivative of $M_{\perp c}(T)$ and $M_{//c}(T)$ curves with respect to temperature (T). This derivative plot is depicted in Fig. 1(b) on a zoomed scale. The 1$^{st}$ derivative of $M_{//c}(T)$ shows a valley at $T_c$ ~714 K while 1$^{st}$ derivative of $M_{\perp c}(T)$ shows a similar valley at $T_c$ ~ 707 K. There is a small difference of 7 K between the $T_c$ obtained from the derivative plots of magnetization $M_{\perp c}(T)$ and $M_{//c}(T)$ curves which, we believe, is mainly due to the slight misorientation of the c-axis with respect to the magnetic field (H) direction and the temperature fluctuations at high temperatures. Since the platelet morphology of the crystal was such that aligning the c-axis parallel to the magnetic field was very accurate, the value obtained by the derivative of $M_{//c}(T)$ has been taken as the correct Curie temperature i.e., $T_c$ ~ 714 K. This $T_c$ is close to the values reported in the earlier single-crystal measurements [62].

It is evident from Fig. 1(a) that with decreasing temperature $M_{\perp c}(T)$ increases upto a peak value for T~ 46K while $M_{//c}(T)$ decreases steadily down to the lowest temperature. The variation of the anisotropy parameter $A = (\frac{M_{\perp c}}{M_{//c}})$ [63] with temperature, depicted in Fig. 1(c), shows a broad peaks at T ~ 46 K suggesting the existence of a transition in the ab-plane. It was found that this peak temperature is sensitive to the measuring field and increases with decreasing field. The values of magnetization measured with field-applied parallel and perpendicular to the c-axis reveal that ~ 28% of the spin component of $M_{//c}(T)$ lies in the ab-plane at T ~ 46 K. These results suggest that the spins are not fully aligned along the c-axis of BFO at low temperatures but may have a significant component transverse to the c-axis in the basal plane (ab-plane) due to the canting of the spins away from the c-axis leading to a magnetic transition at T ~ 46 K for a field of 500 Oe. Such a canting of the spins in BFO has been confirmed recently by X-ray magnetic circular dichroism (XMCD) studies also [55].

**B. Evidence for Two Magnetic Transitions below Room Temperature**

The slope of the $M_{//c}(T)$ curve in Fig. 1(a) was found to change at lower temperatures suggesting that there might be a magnetic transition involving the spin component along the c-axis also. To resolve this transition, we carried out additional magnetization measurements below room temperature using the SQUID based MPMS system. Figs. 2(a) and (b) show the temperature dependence of the dc magnetization on a zero field-cooled single-crystal of BFO along ($M_{//c}$) and perpendicular ($M_{\perp c}$) to the c-axis for $T \leq 300K$ on a highly magnified scale, especially for the $M_{//c}(T)$ plot. These measurements confirm a magnetic transition peak at T~ 46K for the $M_{\perp c}(T)$ plot similar to that observed in Fig 1(c), as shown more clearly in the inset of Fig. 2(b) for measurements under zero field-cooled (ZFC) and field-cooled (FC) protocols. Further, the change of slope at low temperatures in the $M_{//c}(T)$ plot leading to a transition at T~ 25K is also confirmed by these measurements, as can be seen more clearly from the inset of Fig. 2(a). The presence of two transitions along and perpendicular to the c axis is supported by the results of the ac susceptibility ($\chi'(\omega,T)$) measurements carried out along ($\chi'_{//c}(\omega,T)$) and perpendicular ($\chi'_{\perp c}(\omega,T)$) to the c-axis on a zero-field cooled BFO single-crystal, as shown in Fig. 3(a) for a frequency of 700Hz and an ac drive field of 3Oe, using the SQUID based MPMS set-up. Clear signatures of these two transitions were also observed in the $\chi'(T)$ plot of sintered polycrystalline samples of BFO measured using the PPMS system, as shown in Fig. 3(b), for a somewhat higher ac drive field of 13Oe. The transition temperatures obtained from $\chi'_{\perp c}(\omega,T)$ and $\chi'_{//c}(\omega,T)$ plots of Fig. 3(a) are in good agreement with the two peak temperatures in the $\chi'(T)$ plot in Fig. 3(b) for the sintered polycrystalline samples, as illustrated using the two dotted vertical lines passing through the peak temperatures in these figures. The arrows in the figures indicate the transition temperatures. Thus, we conclude that the higher temperature transition at T~ 57 K in Fig. 3(b) for the polycrystalline BFO is related to the transverse component of the spins lying in the ab-plane while the lower temperature transition around 25 K is linked with the longitudinal spin component parallel to the c-axis. It is worth mentioning here that the two anomalies seen in the $\chi'(T)$ plot of Fig. 3(b) for 13 Oe ac drive field are not so well resolved for lower amplitude ac drive fields, especially at low frequencies.

**C. Evidence for Spin-Glass Character of the Two Magnetic Transitions in $\chi'(\omega,T)$**

The irreversibility of the $M_{\perp c}$ plots for the zero-field and field cooled samples with the two curves bifurcating below T~ 80K (see inset of Fig. 2(b)) points towards the possibility of spin-glass character of this transition. Fig. 4 depicts the temperature dependence of the ac-susceptibility ($\chi'(\omega,T)$) of the powder samples of BFO at several frequencies measured using the PPMS system. It is evident from the figure that the two transition temperatures ($T_f(\omega)$) shift towards higher temperature side with increasing measuring frequency (f). Such a shift can occur as a results of blocking transition of superparamagnetic (SPM) clusters [64] or spin-glass freezing [13–15]. The two processes can be distinguished by investigating the temperature dependence of the spin relaxation time ($\tau = 1/\omega$, $\omega = 2\pi f$) which for SPM blocking follows Arrhenius type temperature dependence with linear $\ln(\tau)$ vs $1/T$ plot. The higher temperature $T_f(\omega)$ was obtained by least-squares fit using a $5^{th}$ order polynomial while the lower temperature $T_f(\omega)$ was obtained from the third derivative plot. The non-linear nature of the $\ln(\tau)$ vs $1/T$ plot shown in Figs. 5(a) and (b) for the two peak temperatures rules out SPM blocking. For spin-glass freezing, $\tau$ is known to exhibit critical slowing down with divergence at a characteristics spin-glass transition temperature $T_{SG}$ ($< T_f(\omega)$) below which the ergodic symmetry is broken [13,14]. Such a critical dynamics for spin glasses has been modeled in terms of a power-law behaviour [13,14,65] based on scaling theories [13,14]:

$$\tau = \tau_0((T_f - T_{SG})/T_{SG})^{-z\nu} \qquad \ldots\ldots(2)$$

where $\tau_0$ is the inverse of the attempt frequency and $\nu$ is the critical exponent for the correlation length ($\xi$) while $z$ is the exponent for the power-law dependence of $\xi$ on $\tau$. The least-squares fits for the power-law dynamics are shown by continuous line through the data points in the insets of Figs. 5(a) and (b) with $T_{SG1} \sim (57.5 \pm 0.03)$K, $z\nu_1 \sim (0.82 \pm 0.01)$, $\tau_{01} \sim 3.3 \times 10^{-4}$ s and $T_{SG2} \sim (23.5 \pm 0.03)$K, $z\nu_2 \sim (0.8 \pm 0.04)$ and $\tau_{02} \sim 2.5 \times 10^{-4}$ s for the higher and lower temperature transitions, respectively. Excellent fits seen in Figs. 5(a) and (b) confirm the spin-glass character of the two transitions involving freezing of part of the transverse and longitudinal components of the spins. The larger value of the attempt relaxation time ($\tau_0$) for both the transitions suggests cluster spin-glass behaviour [66,67] as $\tau_0$ is known to be much smaller ($10^{-12}$ to $10^{-13}$ s) for the atomic glasses [13,14,68–70]. Presence of local clusters of spins has been reported in several ordered compounds with frozen-in geometrical frustration by diffuse neutron scattering and small angle neutron scattering (SANS) [71,72] and has formed the basis of several theoretical models of spin-glass transition in ordered compounds [40–43,73–75]. We believe that the large value of $\tau_0$ in BFO is due to the involvement of such spin clusters in the spin-glass freezing while low value of $z\nu$ ($< 1$) indicates unconventional spin-glass behaviour [76].

**C. Further Confirmation of the Spin-Glass Character of the Two Magnetic Transitions**

The two spin-glass freezing temperatures ($T_f(\omega)$) shift to the lower temperatures side under dc bias field (H), as shown in Fig. 6(a) for f = 200Hz. Such a shift in the conventional site-disordered Heisenberg systems follows the well-known de Almeida-Thouless (A-T) and Gabay-Toulouse (G-T) [77–79] lines in the T-H plane as per the general expression:

$$H^m = a[1 - T_f(\omega, H)/T_f(\omega, H = 0)] \qquad \ldots\ldots(3)$$

where m = 2/3 and 2 for A-T and G-T lines, respectively. The least-squares fits to the $T_f(\omega)$ versus H plots for the two transitions shown in Fig. 6(b) give m = (2.0 ± 0.004) and (0.66 ± 0.001) which indeed match with the theoretical exponents for the G-T and A-T lines, respectively, in the conventional spin glasses [80–84].

Below the spin-glass transition temperature $T_{SG}$ (< $T_f(\omega \neq 0, H \neq 0)$), conventional spin glasses exhibit slow relaxation of susceptibility with time (t) [84–86] due to the reorganization of the metastable states of the spin glasses [87,88]. Fig. 7 depicts the evolution of $\chi'(f = 200Hz)$ of BFO as a function of time under dc field bias of 50Oe at 40K (< $T_{SG1}$ = 57.5K) and 10K (< $T_{SG2}$ = 23.5K) on zero-field cooled samples. The least-squares fits to the two plots in Fig. 7 using the well-known Kohlrausch-Williams-Watt (KWW) stretched exponential behaviour [89,90], $\chi'(t) = \chi'_0 + \chi'_g*\exp\{-(t/\tau)^\beta\}$, gives $\chi'_0$ = (0.013 ± 3 x $10^{-7}$) and (0.0199 ± 5 x $10^{-7}$), $\chi'_g$ = -(1.57 x $10^{-4}$ ± 1x$10^{-6}$) and -(1.64 x $10^{-4}$ ± 3 x$10^{-6}$), and β = (0.65 ± 0.003) and (0.66 ± 0.004), respectively, for the two spin-glass transition phases. The fact that the exponent β deviates significantly from β = 1 clearly reveals slow non-Debye type polydispersive nature of $\chi'$ as a function of time under a constant dc field. The values of β for the two spin-glass transitions of BFO are comparable and fall within the expected range 0 < β < 1 for the conventional spin glasses [13,14,86,91].

The relaxation, aging and memory effects are characteristics of the conventional spin glasses as they provide evidence for the evolution of spins from one free-energy minimum to another with time and temperature [87,88,92]. To probe the memory effects in our samples below $T_{SG1}$ and $T_{SG2}$, we used the following protocol [86,93,94]: the sample was initially cooled down to $T_1$ = 40K (< $T_{SG1}$ = 57.5K) under zero magnetic field and $\chi'(t)$ was recorded at 200Hz in the presence of a dc biasing field H = 50Oe over 8 hours [segment "gh" in Fig. 8]. After 8 hours, the sample was quenched to a lower temperature $T_2$ = 35K (such that $T_1$ - $T_2$ = 5K) in the presence of the same biasing field and χ'(t) was again recorded over 6 hours at the same frequency [segment "ij" in Fig. 8]. Finally, the sample was heated back to 40K and then χ'(t) was recorded for 6 hours at the same frequency and dc biasing field [segment "kl" in Fig. 8]. It is evident from Fig. 8 that the relaxation process during the segment "kl" is simply a continuation of the process during the segment "gh". This behavior of χ'(t) suggests that the state of the system before cooling to $T_2$ = 35K is recovered when the sample is heated back to the initial temperature i.e. $T_1$ = 40K. Using the same protocol with ΔT = 5K, we have confirmed memory effects at 10K (< $T_{SG2}$ = 23.5K) for the second spin-glass phase also (see Fig. 8). The observation of non-exponential relaxation and memory effect below $T_{SG1}$ and $T_{SG2}$ confirms the presence of metastable states in the spin-glass phases of BFO, similar to that in the conventional spin glasses [86,94,95].

**D. Signature of Spin Freezing in Neutron Powder Diffraction Studies**

The magnetic contribution to integrated intensity of the Bragg peaks in the neutron diffraction patterns is proportional to the square of the ordered magnetic moment. The temperature dependence of the ordered magnetic moment follows the Brillouin function behaviour:

$$\mu = \mu_0 B_J(x), \qquad \ldots\ldots(4)$$

where $x = \frac{3J}{J+1}\frac{T_c}{T}\frac{\mu}{\mu_0}$. Here J is the total angular momentum, μ and $\mu_0$ are the magnetic moments at a non-zero temperature T and T= 0 K, respectively, and $B_J$ is the Brillouin function, which is given by:

$$B_J(x) = \frac{2J+1}{2J}\coth\left(\frac{2J+1}{2J}x\right) - \frac{1}{2J}\coth(\frac{1}{2J}x), \qquad \ldots\ldots(5)$$

Any disordering of the spin components due to the spin-glass freezing is expected to lead to a decrease in the integrated intensity of reflections having magnetic contributions [84,96]. This in turn is expected to manifest as a departure from the expected square of the Brillouin function behaviour of the magnetic contributions to the integrated intensities [84,96].

Fig. 9 depicts the evolution of the neutron powder diffraction pattern of BFO in the temperature range 10 to 300 K in a limited 2θ range for a few temperatures. Since the unit cell of BFO consists of two formula units, the propagation vector for the ferrimagnetic (FMI) phase corresponds to k = 0 and hence no new peak with only magnetic contribution is expected to be present due to the FMI structure of BFO. The magnetic contribution to the integrated intensity is superimposed over the nuclear contribution, which is largely temperature independent. To assess the magnetic contribution to the nuclear peaks, we refined the room temperature structure of BFO using input structural parameters obtained from Rietveld analysis of the synchrotron X-ray powder diffraction (SXRPD) pattern for the $P6_3/mmc$ space group. Fig. 10(a) depicts the best fit for the nuclear structure model while Fig. 10(b) depicts the difference profile which reveals huge mismatch between the observed and calculated profiles. As confirmed in the next section, this mismatch is due to huge magnetic contributions to several reflections of BFO.

We monitored the temperature dependence of a few *h0l* type reflections such as *103* and *203* with significant magnetic contributions. These reflections were selected as they are stand-alone reflections and therefore their integrated intensities could be obtained very accurately. We did not consider reflections having a partially overlapping peak side by side to avoid any errors in peak deconvolution affecting the integrated intensities. Fig. 11 depicts the temperature dependence of the integrated intensity of two such reflections, *103* and *203* ($I_{103}$ and $I_{203}$). The intensity of these reflections contains information about freezing of both the longitudinal and transverse spin components together. The least-squares fit corresponding to the square of the Brillouin function behaviour, given by Eq. 4 above, is shown using continuous line in Fig. 11. It is evident from this figure that the integrated intensity deviates from the theoretically expected behaviour below 100K. The observed intensities are significantly lower than that expected from the square of the Brilluoin function behaviour. This demonstrates that some parts of the ordered magnetic moments involved in the spin-glass freezing are getting disordered while the remaining parts remain ordered suggesting coexistence of the spin-glass and the LRO ferrimagnetic phases in agreement with the recent XMCD studies [55].

**E. Evidence for Kagome-Spin Configuration for the Basal Plane Component of the Spins**

To determine the evolution of the average spin configuration in BFO as a function of temperature, we carried out Rietveld refinement of the magnetic and nuclear structures using neutron powder diffraction (NPD) patterns recorded at several temperatures with T ≤ 300K. Theoretical calculations of the single-ion anisotropy of the magnetic ions for different sites in BFO suggest that the single-ion anisotropy of $Fe^{3+}$ is negative for the 12k site whereas it is positive for the remaining sites [97,98]. The negative value of single-ion anisotropy indicates that anisotropy energy will be maximum when the magnetic spins at the 12k site are parallel to the c-axis [99]. Mean-field calculations of the exchange interactions in BFO have revealed that magnetic spins at the 12k site experience isotropic exchange interaction [100] as these spins are subjected to competing ferromagnetic interactions with spin at the 2a, 2b, 12k sites (all spin-up) and strong anti-ferromagnetic interaction with the spins at the $4f_1$, $4f_2$ sites (all spin down) [101]. The weak effective exchange interaction [100] and the negative value of anisotropy constant for the 12k site [97,98] may lead to canting of spins at this site. It has been observed experimentally that a very small amount of substitution (3% to 5%) at the $Fe^{3+}$ sites by non-magnetic Al, Ga, Sc, Mn, Zn and In leads to significant canting of the spins at 12k sites [102–107]. All these observations indicate that the spins at the 12k sites are the most probable candidates for canted spin components. To capture the signatures of this, we carried out Rietveld refienement of the magnetic structure using neutrom powder diffraction (NPD) patterns at multiple temperatures below 300K.

For the refinements, we considered all possible Irreps for various magnetic Wyckoff sites in the unit cell using BasIrreps package [57]. The basis vectors of each irrep for different Wyckoff sites are listed in Tables I, II and III. It is evident from these tables that the irrep $\Gamma_2$ is common for all the Wyckoff sites. It is also evident from the tables that this irrep has only one basis vector for all the Wyckoff sites (2a, 2b, 4$f_1$ and 4$f_2$) except the 12k site. For the 12k site, the irrep $\Gamma_2$ has two basis vectors, labeled as $\Gamma_{(2z)}$ and $\Gamma_{(2x)}$, corresponding to the components of the magnetic moment along and perpendicular to the c-axis, respectively. Since the irrep $\Gamma_{(2z)}$ at the 12k site gives rise to finite magnetic moment parallel or antiparallel to the c-axis (see Table I, II and III) for 12k as well as the other Wyckoff sites, it represents the collinear FMI structure of BFO as per the Gorter model [61]. The irrep $\Gamma_{2(x \oplus z)}$, on the other hand, gives rise to components of the moment parallel and perpendicular to the c-axis for the 12k site (see Table III(a) and (b)), while leaving moments at all other Wyckoff sites only along the c-axis. This suggests that the irrep $\Gamma_2$ can give rise to both collinear ($\Gamma_{(2z)}$) and non-collinear ($\Gamma_{2(x \oplus z)}$) FMI structures of BFO.

Since the Gorter model for the magnetic structure of BFO corresponds to a collinear magnetic structure below 714 K, we first carried out refinement for the irrep $\Gamma_{2(z)}$ with only one basis vector. This irrep gives reasonable average fit down to 10 K but the value of the overall $\chi^2$ for this model keeps increasing with decreasing temperature as can be seen from Fig. 12. This behaviour of $\chi^2$ with temperature points towards the possibility of deviation from the collinear model for the magnetic structure of BFO and indicates that the irrep $\Gamma_{2(z)}$ with only one basis vector is not sufficient to explain the real magnetic structure especially at low temperatures. In view of this, we refined the magnetic structure using both the basis vectors of $\Gamma_2$ (i.e., the irrep $\Gamma_{2(x \oplus z)}$. The $\chi^2$-values obtained after refining the magnetic structure at different temperatures using this irrep are, in general, lower than those obtained with only one basis vector, as can be seen from Fig. 12. Further, the $\chi^2$-values for $\Gamma_{2(x \oplus z)}$ model are nearly comparable at 300 K and 250K. However, these values begin to increase below 250K even for the $\Gamma_{2(x \oplus z)}$ model. Evidently, the magnetic structure below 250K requires consideration of other irreps along with $\Gamma_{2(x \oplus z)}$.

Both $\Gamma_{(2z)}$ and $\Gamma_{2(z \oplus x)}$ irreps have a common magnetic space group P6$_3$/mm′c′. To determine the possible combinations of the irreps which can give rise to another non-collinear magnetic structure of BFO below 250K, we considered all possible isotropy subgroups of the space group P6$_3$/mm'c' using the ISODISTORT package [108]. The isotropy magnetic subgroup tree along with the corresponding irrep is shown in Fig. 13. There are eight possible magnetic space groups which are isotropy subgroups of the magnetic space group P6$_3$/mm'c'. Accordingly, we refined the magnetic structure using the irrep combinations ($\Gamma_2 \oplus \Gamma_1$), ($\Gamma_2 \oplus \Gamma_3$), ($\Gamma_2 \oplus \Gamma_4$), ($\Gamma_2 \oplus \Gamma_7$), ($\Gamma_2 \oplus \Gamma_8$), ($\Gamma_2 \oplus \Gamma_9$), ($\Gamma_2 \oplus \Gamma_{10}$) and ($\Gamma_1 \oplus \Gamma_7$) corresponding to the eight magnetic isotropy subgroups P6$_3$/m, P-31c', P-3m'1, P6$_3$m'c', P6$_3$2'2', P-6m'2', P-62'c', and P6$_3$22 one by one. A perusal of the basis functions of $\Gamma_{2(x \oplus z)}$ as well as the eight possible isotropy subgroups with different irrep combinations shows that these space groups lead to canting of the magnetic spins at the 12k Wyckoff site only. We find that the refinements of the magnetic structure using irrep combinations ($\Gamma_2 \oplus \Gamma_7$), ($\Gamma_2 \oplus \Gamma_8$), ($\Gamma_2 \oplus \Gamma_9$), ($\Gamma_2 \oplus \Gamma_{10}$) and ($\Gamma_1 \oplus \Gamma_7$) give rise to very large $\chi^2$-values in comparison to the values obtained for ($\Gamma_2 \oplus \Gamma_1$), ($\Gamma_2 \oplus \Gamma_3$) and ($\Gamma_2 \oplus \Gamma_4$) combinations. From the refined structure, we also find that only the $\Gamma_2 \oplus \Gamma_3$ and $\Gamma_2 \oplus \Gamma_4$ irrep combinations give rise to a significant reduction in the $\chi^2$ and magnetic agreement factor R$_M$ values with respect to the $\Gamma_{2(x \oplus z)}$ model, as shown in Fig. 12 at T < 250 K (i.e., 100 K and 50 K in the figure). Since the number of refinable parameters for $\Gamma_{2(x \oplus z)} \oplus \Gamma_3$ (magnetic space group $P\bar{3}1c'$) and $\Gamma_{2(x \oplus z)} \oplus \Gamma_4$ (magnetic space group $P\bar{3}m'1$) are 3 and 4, respectively, we choose $\Gamma_{2(x \oplus z)} \oplus \Gamma_3$ combination as it gives comparable fit for one less refinable parameter. On lowering the temperature to 10 K, the $\Gamma_{2(x \oplus z)} \oplus \Gamma_4$ gives slightly lower $\chi^2$ and $R_M$ values than those for $\Gamma_{2(x \oplus z)} \oplus \Gamma_3$. However, the difference is not significant enough to choose one or the other combination of irreps keeping in mind that $\Gamma_{2(x \oplus z)} \oplus \Gamma_4$ combination has one additional refinable parameter. We therefore believe that the magnetic structure of BFO at T ≲ 100 K corresponds to the $\Gamma_{2(x \oplus z)} \oplus \Gamma_3$ combination of irreps. The fits between the observed and calculated profiles, obtained after Rietveld refinement of the magnetic and nuclear structures using irrep $\Gamma_{2(x \oplus z)}$ and $\Gamma_{2(x \oplus z)} \oplus \Gamma_3$ at 300 K and 50 K, respectively, are shown in Figs. 14(a) and 15(a), respectively. All these fits are quite satisfactory. We also depict the calculated nuclear and magnetic contributions to the diffraction profile obtained by Rietveld refinement in Figs. 14(b) and (c) and Figs. 15(b) and (c) for 300K and 50K, respectively. A comparison of Figs. 14(b) and (c) with those shown in Figs. 10(a) and (b), however, confirm large magnetic contributions to several nuclear peaks.

The $\Gamma_{2(x \oplus z)} \oplus \Gamma_3$ combination of irreps with $P\bar{3}1c'$ magnetic space group of BFO leads to kagome spin configuration for the 12k Wyckoff sites. This is shown schematically in Fig. 16 for the transverse spin components in the basal plane. This spin configuration is fully frustrated with macroscopic degeneracy [109]. The longitudinal component of the spins for the 12k site as well as the spins at the 2a, 2b, $4f_1$, and $4f_2$ Wyckoff sites remain aligned along the c-axis for the $P\bar{3}1c'$ magnetic space group also, as per the Gorter model. Thus, the main difference between the magnetic structure of BFO proposed here and that proposed earlier by Gorter [61] is due to the canting of spins at the 12k Wyckoff sites (Fig. 16(b)) which gives rise to kagome spin configuration for the component of the spins in the ab- plane. Both LRO AFM and spin liquid behaviours [110] have been reported in kagome spin configurations depending on the strength of the interlayer coupling [111]. Stronger z-coupling favours LRO state. In our case, the interplanar coupling between the kagome layers is very strong, as only a component of the spins, that too at only the 12k Wyckoff site, lies in the basal plane while the spins at all other Wyckoff sites remain aligned in the z-direction. So our system is far from being a quasi-two dimensional kagome type. The strong out of plane coupling leads to the average spin configurations shown in Figs. 16(a) and (b).

We believe that the genesis of the spin-glass transition in BFO is the emergence of the highly degenerate kagome spin configuration for the basal plane component of the spins at the 12k site as a result of change in the magnetic space group from $P6_3/mm'c'$ to $P\bar{3}1c'$. In the $P\bar{3}1c'$ magnetic space group of BFO, the bond lengths are not equal in the basal plane or in the c-direction (see Table IV) and this may be responsible for the randomness in the exchange interactions which in conjunction with geometrical frustration can give rise to spin-glass transitions [40–42,73,74]. The shortest linkage between the successive kagome layers, say at z = -0.10855 and 0.10855, as shown in Fig. 17, in the perpendicular direction is via the spins at the 2a site. This linkage gives rise to a pyrochlore slab of two inverted tetrahedra formed by the spins at the 12k site in the ab plane and the spins at the 2a site, as shown schematically in Fig. 17. The shortest linkage between the successive kagome bilayer spin configurations (12k-2a-12k) in BFO and SCGO is via the spins at the $4f_2$ Wyckoff sites at z= 0.1903 and 0.3096, as shown in Fig. 17. This structure is reminiscent of the magnetic structure of the well-known geometrically frustrated $SrCr_xGa_{12-x}O_{19}$ (SCGO) system [17,33,112–114] which exhibits non-conventional spin-glass transition for $5 \leq x \leq 8.78$ [17,34,35,114]. A brief comparison of the low temperature behaviour of BFO with that of the SCGO system would, therefore, be in order as it underlines the novelty of our findings in the field of geometrically frustrated magnets: (1) Unlike the SCGO, where there is only partial coverage of the magnetic sublattice by the $Cr^{3+}$ spins [115], BFO is free of any substitutional disorder with 100% coverage by $Fe^{3+}$ spins. For example, ~14% and ~4% of $Cr^{3+}$ at the 12k and 2a sites, respectively, are occupied by non-magnetic $Ga^{3+}$ in SCGO, [115] whereas both the sites are fully occupied by $Fe^{3+}$ in BFO. Similarly, the occupancy of $Ga^{3+}$ at the $4f_2$ site is ~13.5% in SCGO whereas it is 100% $Fe^{3+}$ for BFO. (2) The spins of SCGO lie in the basal plane [33,34] whereas the BFO spins have components both along the c-axis as well as in the basal plane. (3) Unlike SCGO, where there is no LRO transition preceding the spin-glass transition [17], BFO shows spin-glass transition due to freezing of the transverse and some part of longitudinal of components of the spins of the LRO ferrimagnetic phase which continues to coexist with spin-glass phases. (4) The geometrical frustration of BFO is emergent in nature and appears in the LRO ferrimagnetic phase of BFO at low temperatures whereas it is frozen-in in the crystal structure of the paramagnetic phase of SCGO [17]. (5) Most significantly, the spin-glass transitions in BFO mimic the theoretical predictions for the disordered Heisenberg systems near percolation threshold even though there is no apparent chemical site-disorder, whereas SCGO shows non-conventional spin-glass behaviour [2,35]. We believe that our findings will stimulate theoretical studies to unravel the physics of spin-glass transitions in LRO systems due to emergent geometrical frustration, an aspect that has remained unexplored so far. We also believe that this work will encourage further experimental studies in search of low temperature spin-glass transition in LRO phases of various hexaferrites [49] and even other LRO magnetic compounds with spins arranged on the triangular, pyrochlore and spinels lattices.

## 4. Summary

We presented evidence for two magnetic transitions involving transverse and longitudinal spin components using M(T) and $\chi(\omega,T)$ studies on single crystalline and powder samples of $BFe_{12}O_{19}$. Their spin-glass character has been confirmed by the divergence of the spin relaxation time with spin-glass transition temperatures $T_{SG}$ = 57.5K and 23.5K for the transverse and longitudinal components of the spins, respectively. The existence of the two spin-glass transition is supported by the field (H) dependent shift of the spin-glass freezing temperature $T_f(H)$ along the G-T and A-T lines for the transverse and longitudinal components of the spins. The existence of two spin-glass phases has been confirmed by the observation of non-exponential relaxation of the isothermal remanent magnetization as well as memory effects below the two $T_{SG}$. Powder neutron diffraction studies reveal anomalous reduction in the integrated intensity of the *h0l* type Bragg peaks at T ≲100K, suggesting that some parts of the spins begin to get disordered below this temperature. Evidence for the emergence of kagome spin configurations with macroscopic degeneracy at T ≲100 K was presented using Rietveld analysis of the temperature-dependent NPD data. Rietveld refinement gives only the average spin configurations. In real systems, the spins would show disorder from the average kagome spin configuration due to the frustrated exchange interactions. Such a disorder is expected to lead to magnetic diffuse scattering in single-crystal diffraction patterns using polarised neutrons [116–120]. The anisotropic nature of exchange interactions is argued to provide randomness to the few body exchange interactions, which in conjunction with the fully frustrated emergent kagome spin configuration at T ≲100K, may be responsible for the spin-glass freezing in BFO as per the existing theories of the spin-glass transitions in geometrically frustrated ordered compounds [40–43,73,74]. However, there is no theory of spin-glass transition in geometrically frustrated ordered compounds that predicts a succession of transverse and longitudinal spin-glass freezing in coexistence with the LRO phase as a result of emergent geometrical frustration. The current theories are based on pre-existing geometrical frustration due to the spins arranged on specific lattice geometries in the paramagnetic state itself. We believe that our observations would stimulate further theoretical studies to understand the effect of emergent geometrical frustration in LRO phases as a function of temperature.

## Figure Captions

**Figure 01:** (a) Temperature dependence of the dc magnetization $M_{\perp c}$ and $M_{//c}$ measured with a field of 500 Oe applied parallel (//) and perpendicular ($\perp$) to the c-axis of a zero-field cooled (ZFC) single-crystal using the PPMS set-up. (b) The first derivative plot of $M_{\perp c}$ and $M_{//c}$ with respect to the temperature. (c) Variation of the magnetic anisotropy parameter $A = (\frac{M_{\perp c}}{M_{//c}})$ with temperature.

**Figure 02:** Temperature dependence of the dc magnetization (M) measured (a) parallel (//) and (b) perpendicular ($\perp$) to the c-axis of a zero-field cooled (ZFC) single-crystal using the MPMS set-up. Inset of (a) depicts the variation of $M_{//c}$ with temperature on a zoomed scale showing an anomaly around 25K. Inset of (b) depicts the variation of transverse magnetization ($M_{\perp c}$) with temperature under the ZFC and FC protocols showing history dependent irreversibility. The continuous solid lines in these plots are guide to the eyes.

**Figure 03:** (a) Temperature dependence of the real part of the ac susceptibility ($\chi'$) measured with the ac drive field applied perpendicular ($\perp$) and parallel (//) to the c-axis of a zero-field cooled single-crystal using the MPMS set-up. (b) The temperature dependence of the real part of ac susceptibility of zero-field cooled powder sample measured using the PPMS set-up. Continuous solid lines in (a) are guide to the eyes. The arrows indicate the transition temperatures in (a) and (b). The dotted vertical line has been drawn to illustrate that the two peak temperatures $\chi'(T)$ in (b) are close to the peak temperatures in the longitudinal $\chi'_{//c}(T)$ and transverse $\chi'_{\perp c}(T)$ plots shown in (a).

**Figure 04:** Temperature dependence of the real part of ac susceptibility $\chi'(\omega, T)$ of the powder sample at various frequencies using the PPMS set-up. Successive curves are shifted vertically by 0.00015emu/gOe for clarity. The solid lines are guide to the eyes.

**Figure 05:** The $\ln(\tau)$ versus temperature plot showing non-Arrhenius behaviour for the (a) transverse and (b) longitudinal spin-glass transitions. The continuous lines through the data points in the insets of (a) and (b) depict the least-squares fits for the power-law type temperature dependence of relaxation time ($\tau$).

**Figure 06:** (a) Evolution of χ'(T) measured at f = 200Hz of powder sample under different dc biasing magnetic field. The solid lines (blue coloured) though the data points are guide to the eyes. (b) The variation of spin-glass freezing temperature $T_f$ with magnetic field along with the least-squares fits for the Gabay-Toulouse (G-T) and de Almeida-Thouless (A-T) lines through the data points in the T-H plane with m ≃ 2 and ≃ 0.66, respectively.

**Figure 07:** Evolution of the iso-thermal remanent ac susceptibility χ' with time (t), measured at f = 200Hz under 50 Oe dc biasing field at 40K and 10K. The solid lines though the data points are the least-squares fit to KWW equation.

**Figure 08:** Evolution of the iso-thermal remanent ac susceptibility χ' with time (t), measured at f = 200Hz under 50 Oe dc biasing field at 40K and 10K with an intermediate quenching to 35K and 5K, respectively. The solid lines though the data points are guide to the eyes.

**Figure 09:** Powder neutron diffraction pattern of $BaFe_{12}O_{19}$ samples recorded at various temperatures. Successive curves are shifted vertically by a constant value 10 for clarity.

**Figure 10:** (a) Observed (filled red circles), calculated (continuous black line) and (b) difference (green line) profiles obtained from the Rietveld refinement of the nuclear structure using neutron diffraction data at 300K. The vertical tick marks (blue) in (a) represent the Bragg peak positions. Panel (b) depicts the difference profile highlighting the huge magnetic contribution.

**Figure 11:** Temperature dependence of the integrated intensity of (a) *103* and (b) *203* Bragg reflections in the neutron diffraction patterns shown in Fig 10. Broken line is guide to the eyes while the continuous solid black line is the fit for the square of Brillouin function behaviour of the magnetic moment μ.

**Figure 12:** Agreements factors for Rietveld refinement of the nuclear and magnetic structures for various irreps using the neutron powder diffraction patterns collected at different temperatures: (a) overall $\chi^2$ and (b) magnetic agreement factor $R_M$.

**Figure 13:** Isotropy subgroup tree for the magnetic space group $P6_3/mm'c'$

**Figure 14:** (a) Observed (filled red circles), calculated (continuous black line), and difference (bottom green line) profiles obtained after Rietveld refinement of the nuclear and magnetic

structures at 300K for the irrep $\Gamma_{2(x \oplus z)}$. The vertical tick marks above the difference profile represent the Bragg peak positions. The nuclear and magnetic contributions to the calculated profiles are given in panels (b) and (c), respectively. Note the similarity of the intensity profile in (c) with that in Fig. 10 (b).

**Figure 15:** (a) Observed (filled red circles), calculated (continuous black line), and difference (bottom green line) profiles obtained after Rietveld refinement of the nuclear and magnetic structures at 50K for the irreps $\Gamma_{2(x \oplus z)} \oplus \Gamma_3$. The vertical tick marks above the difference profile represent the Bragg peak positions. The nuclear and magnetic contributions to the calculated profiles are given in panels (b) and (c), respectively.

**Figure 16:** Schematic depiction of magnetic spin configurations obtained from Rietveld refinement of the nuclear and magnetic structures at 50K using irrep $\Gamma_{2(x \oplus z)} \oplus \Gamma_3$ (a) longitudinal components for all the Wyckoff sites and (b) transverse component for the 12k Wyckoff site.

**Figure 17:** Schematic depiction of kagome bilayer configuration formed by basal plane spin components at the 12k Wyckoff site linked via spins at the 2a site giving rise to a pyrochlore slab. The linkage between the successive bilayers (12k-2a-12k) via $4f_2$ sites is also depicted in the same figure.

**Figures**

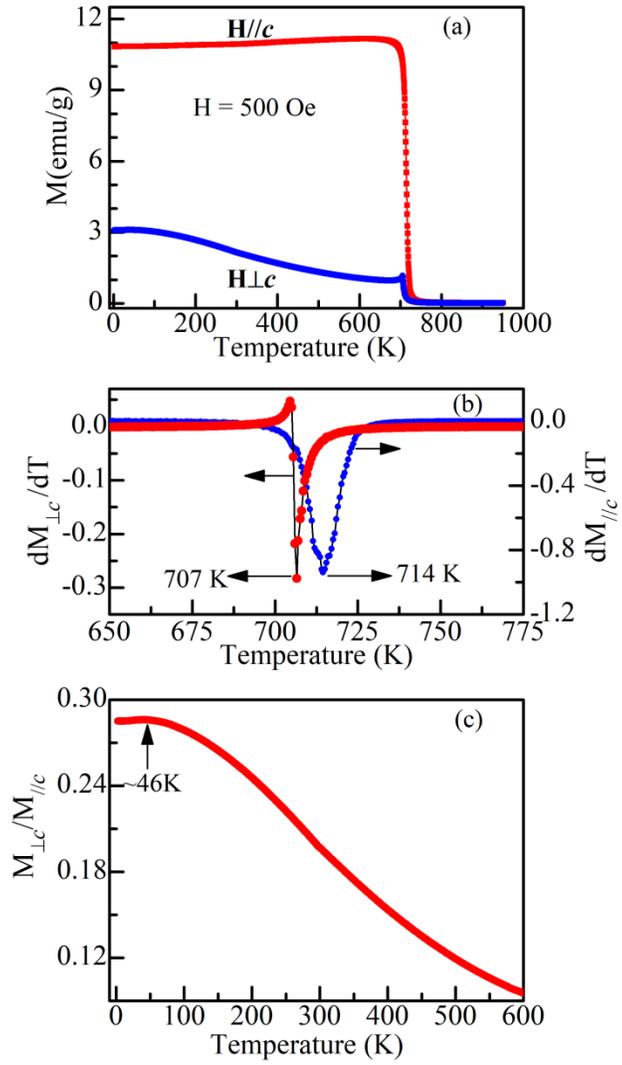

**Figure 01**

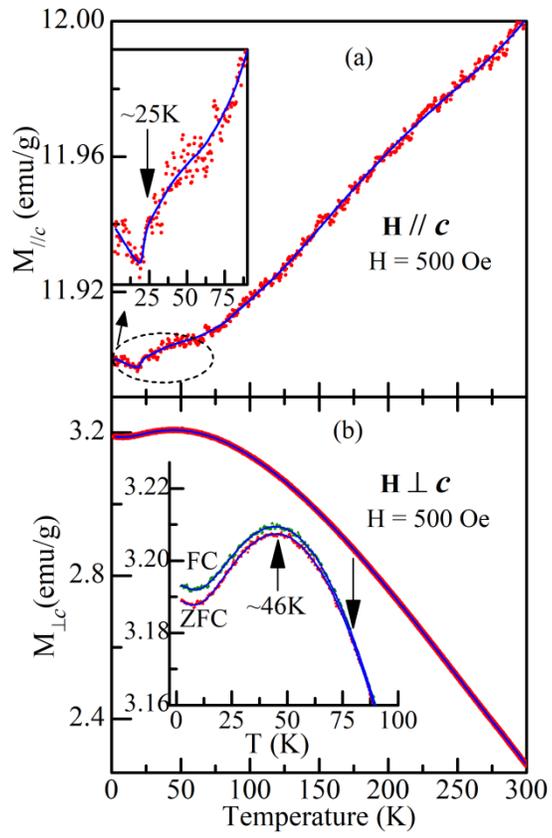

**Figure 02**

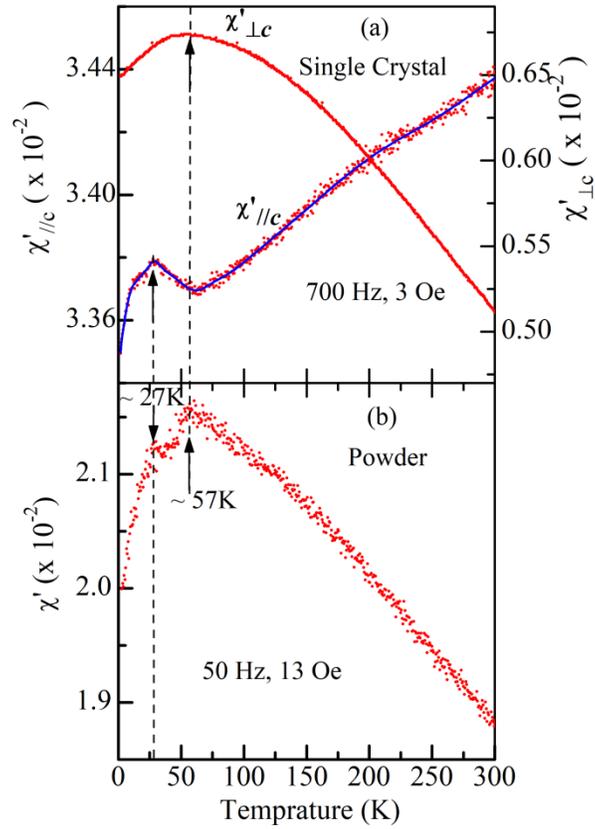

**Figure 03**

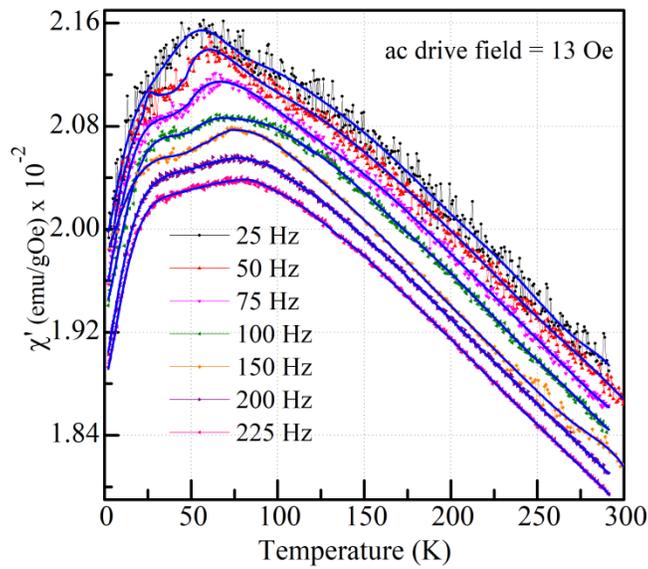

**Figure 04**

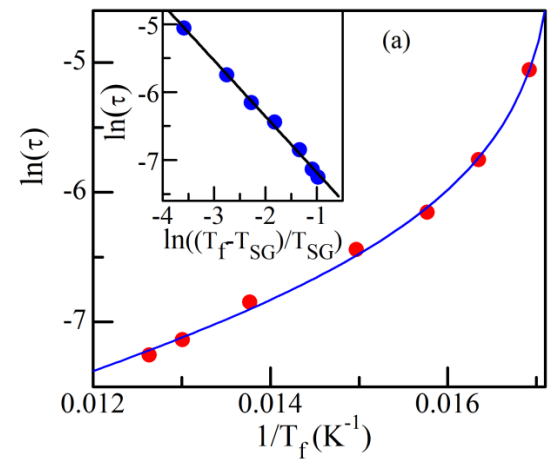

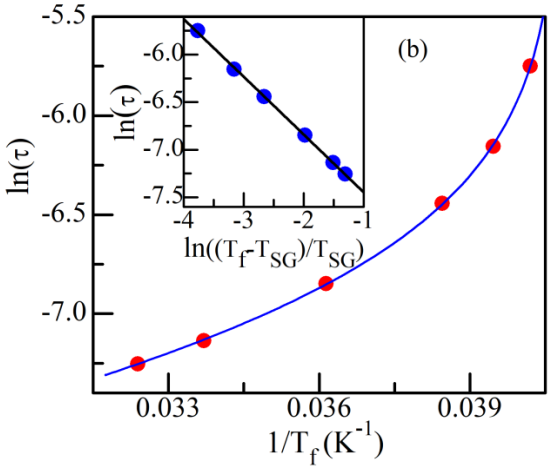

**Figure 05**

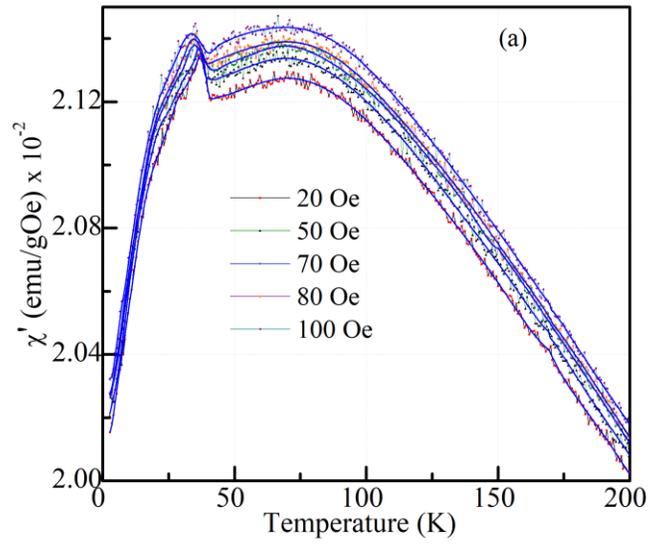

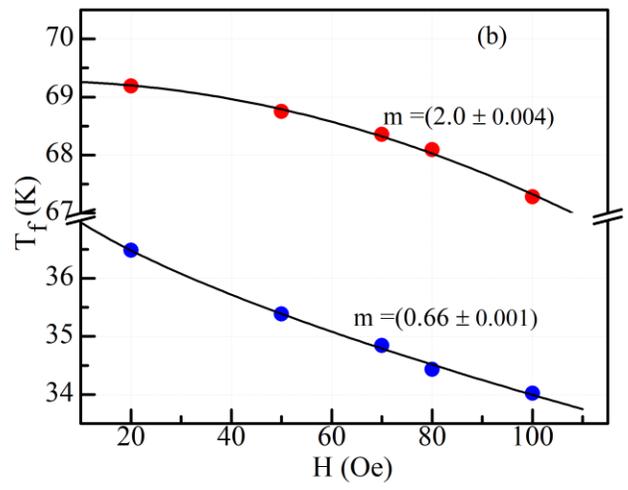

**Figure 06**

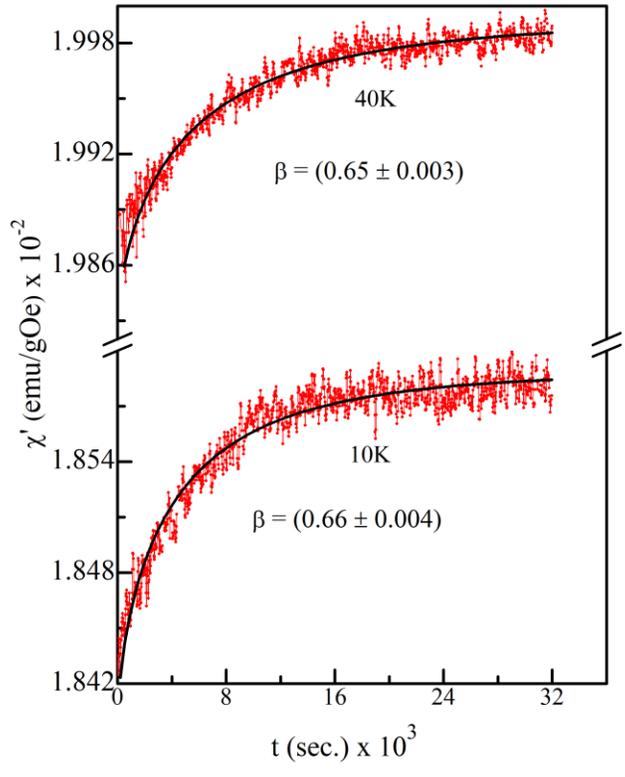

**Figure 07**

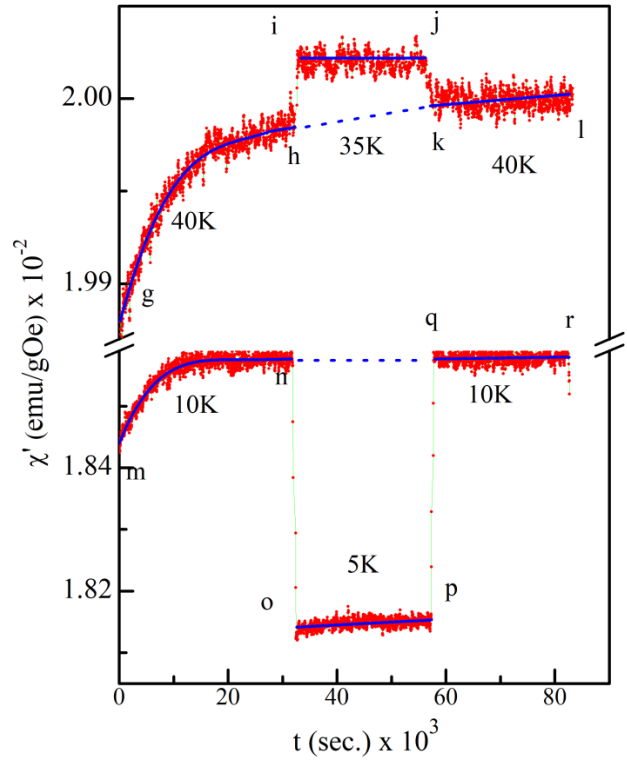

**Figure 08**

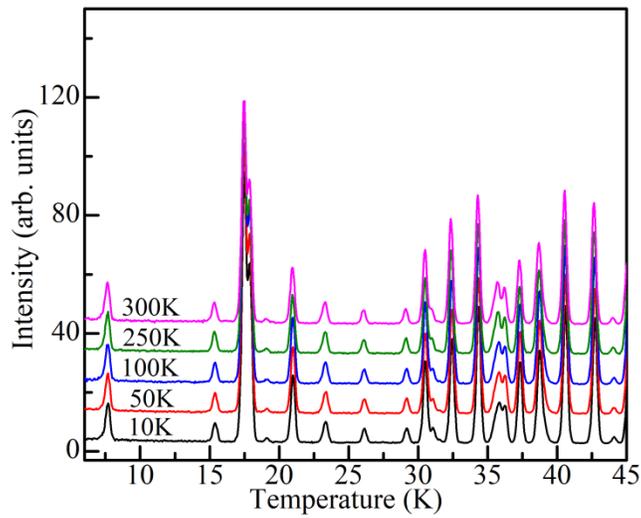

**Figure 09**

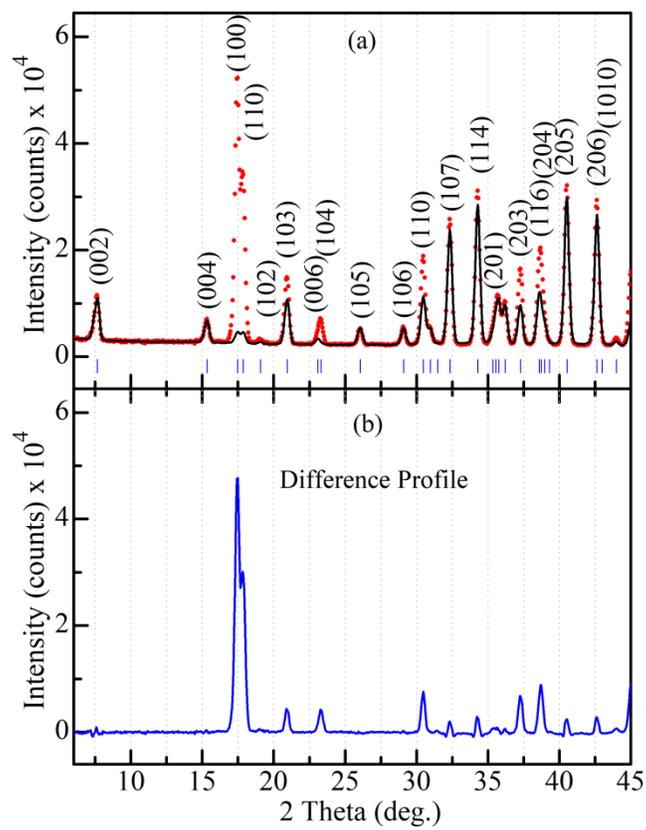

**Figure 10**

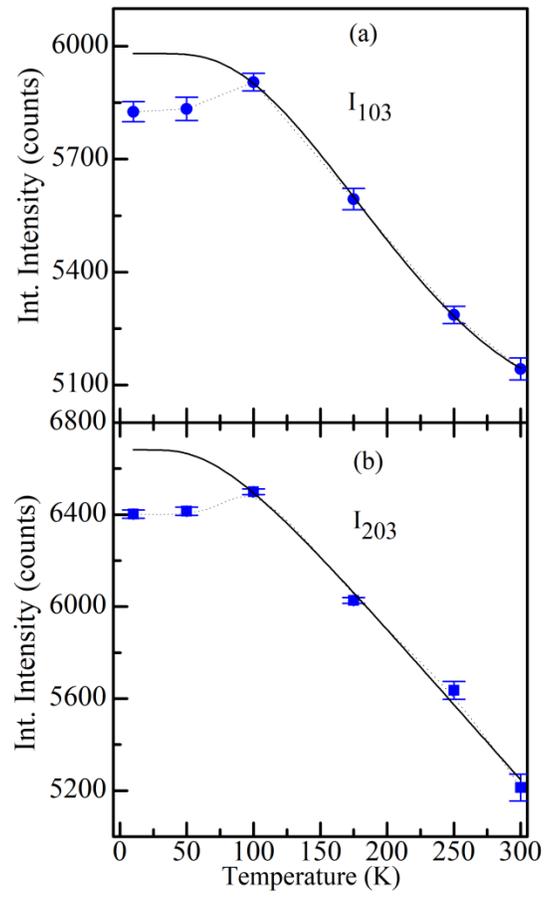

**Figure 11**

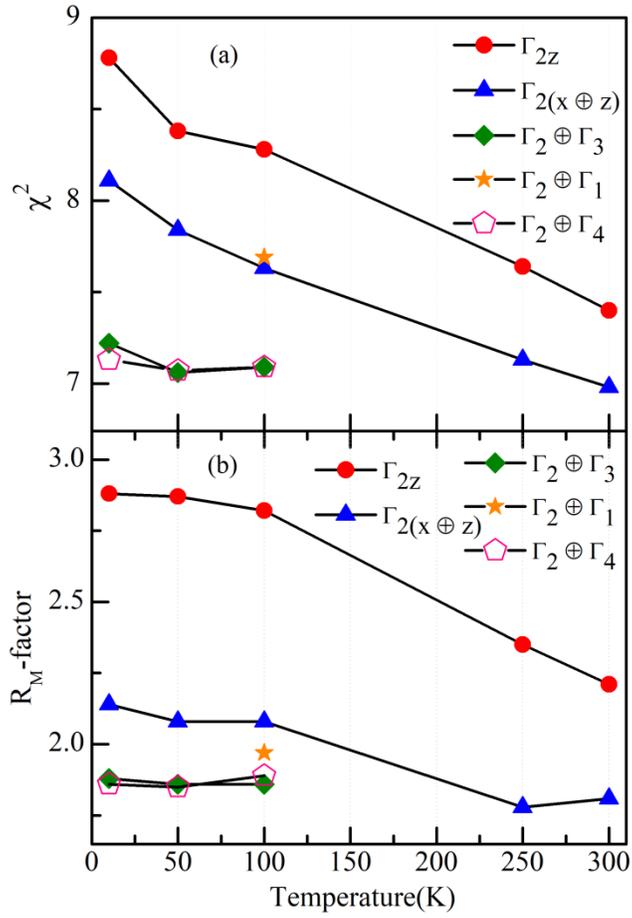

**Figure 12**

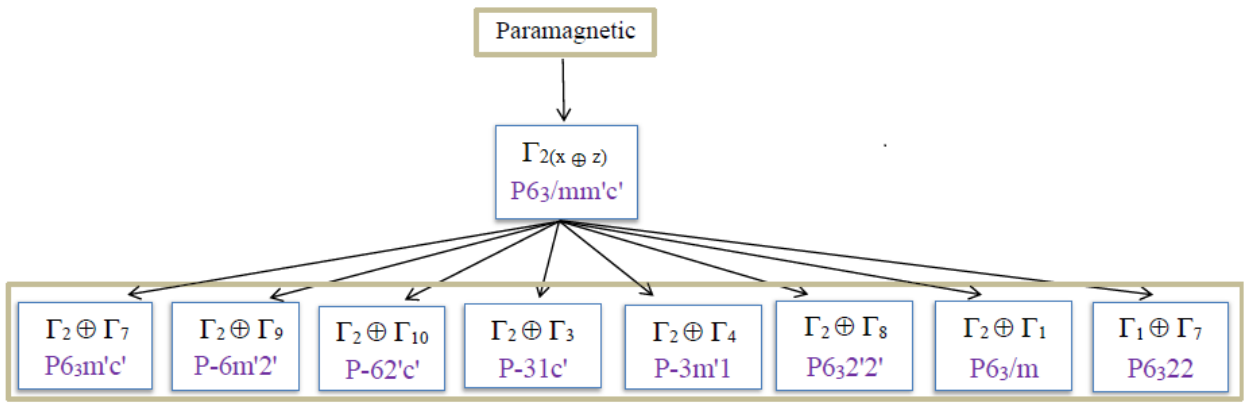

**Figure 13**

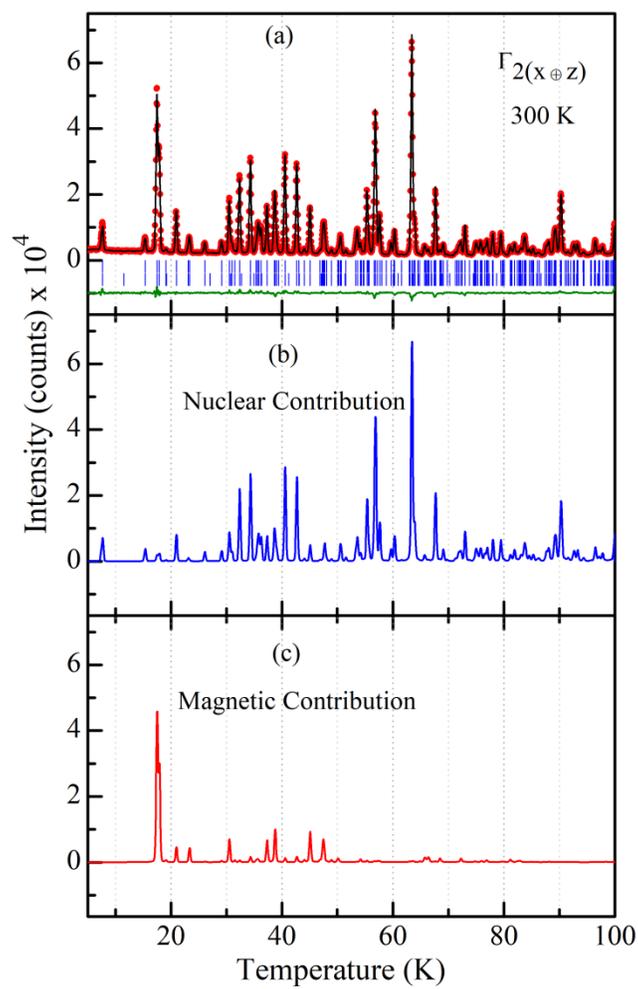

**Figure 14**

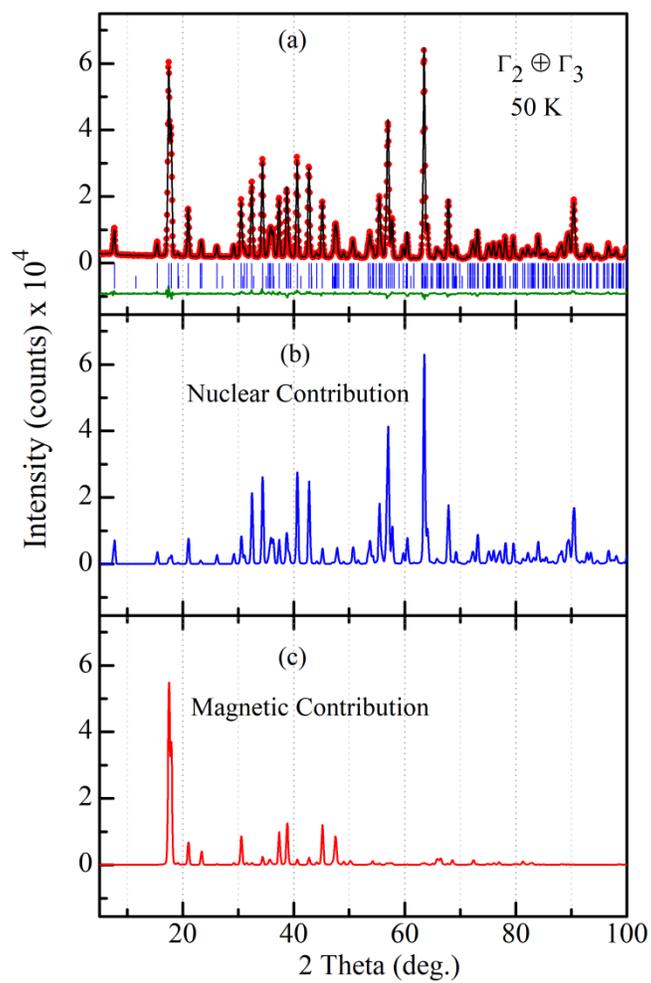

**Figure 15**

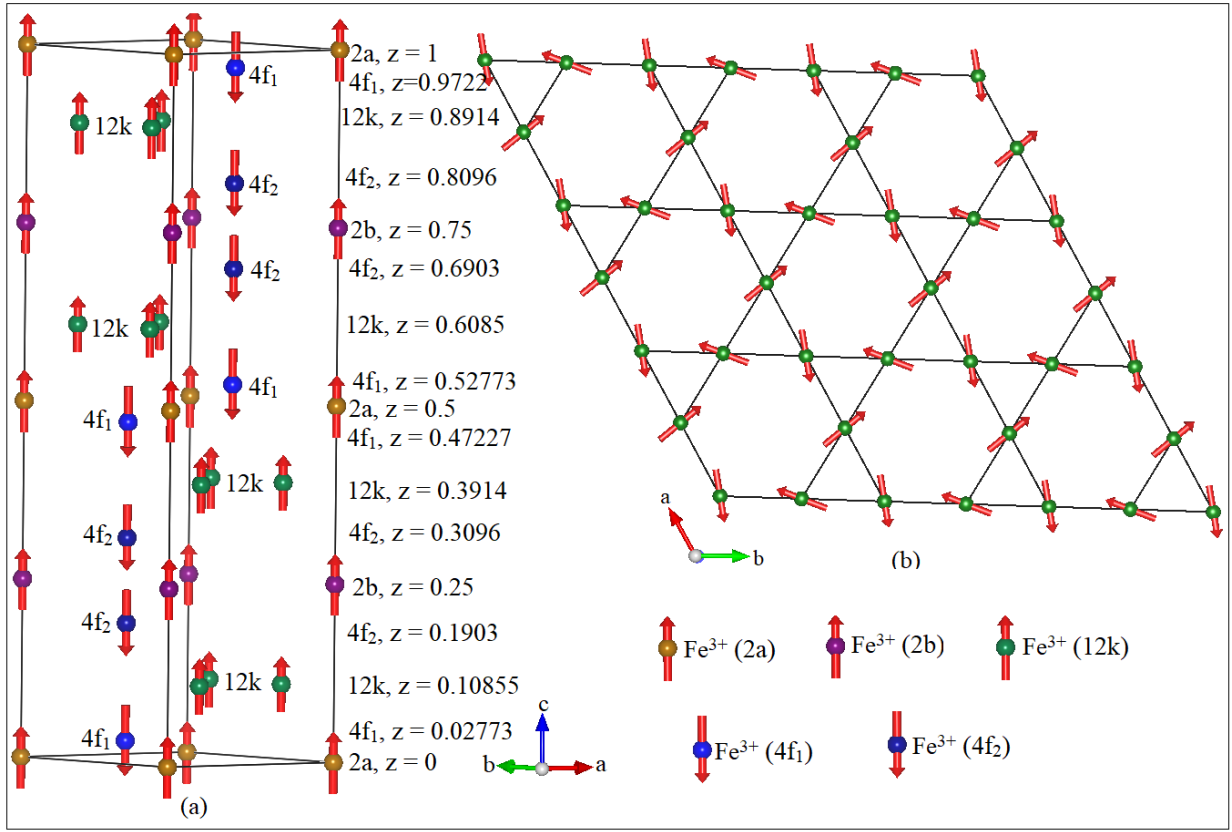

**Figure 16**

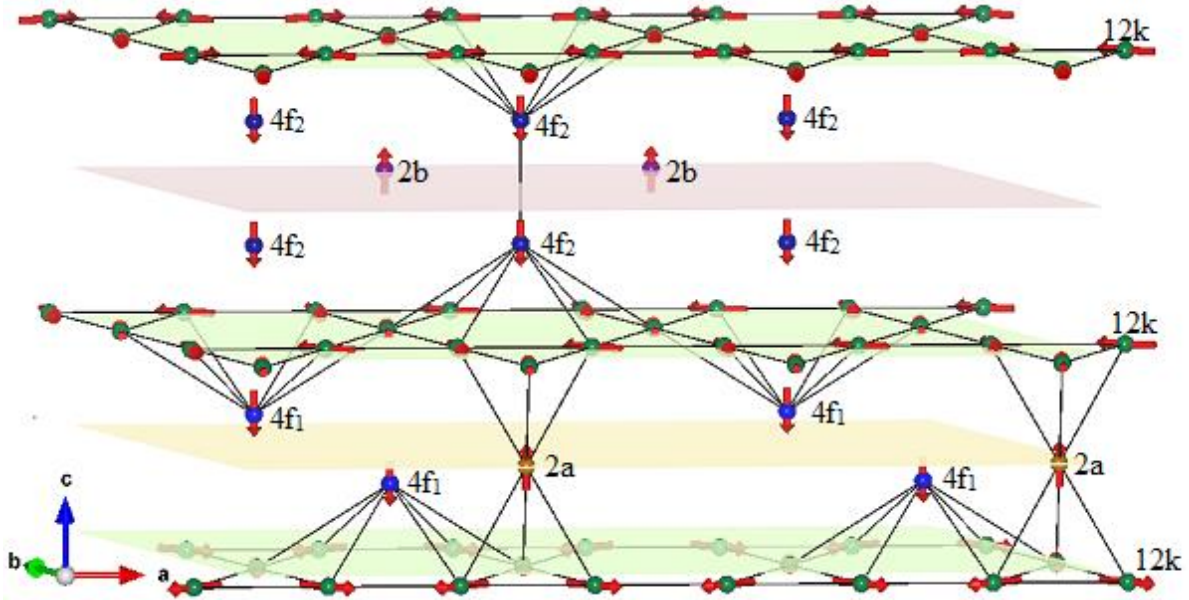

**Figure 17**

# Tables

Table I: Basis vectors of the irreducible representation $\Gamma_n$ for magnetic ion ($Fe^{3+}$) named as Fe1 at 2a Wyckoff site with fractional coordinate (x=0, y=0, z=0) and Fe2 at 2(b) Wyckoff site with fractional coordinate (x=0, y=0, z=0.25).

| Magnetic Representation | Fe1 (2a) $\begin{pmatrix} x \\ y \\ z \end{pmatrix}$ | $\begin{pmatrix} x \\ y \\ z+1/2 \end{pmatrix}$ | Magnetic Representation | Fe2 (2b) $\begin{pmatrix} x \\ y \\ z \end{pmatrix}$ | $\begin{pmatrix} x \\ y \\ z+1/2 \end{pmatrix}$ |
|---|---|---|---|---|---|
| $\Gamma_2$ | $\begin{pmatrix} 0 \\ 0 \\ 1 \end{pmatrix}$ | $\begin{pmatrix} 0 \\ 0 \\ 1 \end{pmatrix}$ | $\Gamma_2$ | $\begin{pmatrix} 1 \\ 0 \\ 0 \end{pmatrix}$ | $\begin{pmatrix} 1 \\ 0 \\ 0 \end{pmatrix}$ |
| $\Gamma_4$ | $\begin{pmatrix} 0 \\ 0 \\ 1 \end{pmatrix}$ | $\begin{pmatrix} 0 \\ 0 \\ -1 \end{pmatrix}$ | $\Gamma_6$ | $\begin{pmatrix} 1 & -1 \\ 2 & 0 \\ 0 & 0 \end{pmatrix}$ | $\begin{pmatrix} 1 & -1 \\ 2 & 0 \\ 0 & 0 \end{pmatrix}$ |
| $\Gamma_5$ | $\begin{pmatrix} 1 & 1 \\ 0 & 2 \\ 0 & 0 \end{pmatrix}$ | $\begin{pmatrix} -1 & -1 \\ 0 & -2 \\ 0 & 0 \end{pmatrix}$ | $\Gamma_9$ | $\begin{pmatrix} 1 \\ 0 \\ 0 \end{pmatrix}$ | $\begin{pmatrix} -1 \\ 0 \\ 0 \end{pmatrix}$ |

Table II: Basis vectors of the irreducible representation $\Gamma_n$ for magnetic ion Fe3 and Fe4 at $4f_{iv}$ (4f1) site with fractional coordinate (x=1/3, y=2/3, z=0.0272) and at $4f_{vi}$ (4f2) site with fractional coordinate (x=1/3, y=2/3, z=0.1904), respectively.

| Magnetic Representation | Fe3 ($4f_1$) and Fe4 ($4f_2$) $\begin{pmatrix} x \\ y \\ z \end{pmatrix}$ | $\begin{pmatrix} -x \\ -y \\ z+1/2 \end{pmatrix}$ | $\begin{pmatrix} x-y \\ -y \\ -z \end{pmatrix}$ | $\begin{pmatrix} x \\ x-y \\ -z+1/2 \end{pmatrix}$ |
|---|---|---|---|---|
| $\Gamma_2$ | $\begin{pmatrix} 0 \\ 0 \\ 1 \end{pmatrix}$ | $\begin{pmatrix} 0 \\ 0 \\ 1 \end{pmatrix}$ | $\begin{pmatrix} 0 \\ 0 \\ 1 \end{pmatrix}$ | $\begin{pmatrix} 0 \\ 0 \\ 1 \end{pmatrix}$ |
| $\Gamma_4$ | $\begin{pmatrix} 0 \\ 0 \\ 1 \end{pmatrix}$ | $\begin{pmatrix} 0 \\ 0 \\ -1 \end{pmatrix}$ | $\begin{pmatrix} 0 \\ 0 \\ 1 \end{pmatrix}$ | $\begin{pmatrix} 0 \\ 0 \\ -1 \end{pmatrix}$ |
| $\Gamma_5$ | $\begin{pmatrix} 1 & 1 \\ 0 & 2 \\ 0 & 0 \end{pmatrix}$ | $\begin{pmatrix} -1 & -1 \\ 0 & -2 \\ 0 & 0 \end{pmatrix}$ | $\begin{pmatrix} 1 & 1 \\ 0 & 2 \\ 0 & 0 \end{pmatrix}$ | $\begin{pmatrix} -1 & -1 \\ 0 & -2 \\ 0 & 0 \end{pmatrix}$ |
| $\Gamma_6$ | $\begin{pmatrix} 1 & -1 \\ 2 & 0 \\ 0 & 0 \end{pmatrix}$ | $\begin{pmatrix} 1 & -1 \\ 2 & 0 \\ 0 & 0 \end{pmatrix}$ | $\begin{pmatrix} 1 & -1 \\ 2 & 0 \\ 0 & 0 \end{pmatrix}$ | $\begin{pmatrix} 1 & -1 \\ 2 & 0 \\ 0 & 0 \end{pmatrix}$ |
| $\Gamma_7$ | $\begin{pmatrix} 0 \\ 0 \\ 1 \end{pmatrix}$ | $\begin{pmatrix} 0 \\ 0 \\ 1 \end{pmatrix}$ | $\begin{pmatrix} 0 \\ 0 \\ -1 \end{pmatrix}$ | $\begin{pmatrix} 0 \\ 0 \\ -1 \end{pmatrix}$ |
| $\Gamma_9$ | $\begin{pmatrix} 0 \\ 0 \\ 1 \end{pmatrix}$ | $\begin{pmatrix} 0 \\ 0 \\ -1 \end{pmatrix}$ | $\begin{pmatrix} 0 \\ 0 \\ -1 \end{pmatrix}$ | $\begin{pmatrix} 0 \\ 0 \\ 1 \end{pmatrix}$ |
| $\Gamma_{11}$ | $\begin{pmatrix} 2 & 0 \\ 1 & -1 \\ 0 & 0 \end{pmatrix}$ | $\begin{pmatrix} -2 & 0 \\ -1 & 1 \\ 0 & 0 \end{pmatrix}$ | $\begin{pmatrix} -2 & 0 \\ -1 & 1 \\ 0 & 0 \end{pmatrix}$ | $\begin{pmatrix} 2 & 0 \\ 1 & -1 \\ 0 & 0 \end{pmatrix}$ |
| $\Gamma_{12}$ | $\begin{pmatrix} 1 & 1 \\ 0 & 2 \\ 0 & 0 \end{pmatrix}$ | $\begin{pmatrix} 1 & 1 \\ 0 & 2 \\ 0 & 0 \end{pmatrix}$ | $\begin{pmatrix} -1 & -1 \\ 0 & -2 \\ 0 & 0 \end{pmatrix}$ | $\begin{pmatrix} -1 & -1 \\ 0 & -2 \\ 0 & 0 \end{pmatrix}$ |

Table III(a): Basis vectors of the irreducible representation $\Gamma_n$ for 6 out of 12 equivalent positions of Fe5 at 12k site with fractional coordinate (x=0.168, y=2x, z=-0.1082).

| Magnetic Representation | Fe5 (12k) | | | | | |
|---|---|---|---|---|---|---|
| | $\begin{pmatrix} x \\ y \\ z \end{pmatrix}$ | $\begin{pmatrix} -x \\ -y \\ z-1/2 \end{pmatrix}$ | $\begin{pmatrix} -y \\ x-y \\ z \end{pmatrix}$ | $\begin{pmatrix} -x+y \\ -x \\ z \end{pmatrix}$ | $\begin{pmatrix} x-y \\ x \\ z-1/2 \end{pmatrix}$ | $\begin{pmatrix} y \\ -x+y \\ z-1/2 \end{pmatrix}$ |
| $\Gamma_1$ | $\begin{pmatrix} 1 \\ 0 \\ 0 \end{pmatrix}$ | $\begin{pmatrix} -1 \\ 0 \\ 0 \end{pmatrix}$ | $\begin{pmatrix} 0 \\ 1 \\ 0 \end{pmatrix}$ | $\begin{pmatrix} -1 \\ -1 \\ 0 \end{pmatrix}$ | $\begin{pmatrix} 1 \\ 1 \\ 0 \end{pmatrix}$ | $\begin{pmatrix} 0 \\ -1 \\ 0 \end{pmatrix}$ |
| $\Gamma_2$ | $\begin{pmatrix} 0 & 1 \\ 0 & 2 \\ 1 & 0 \end{pmatrix}$ | $\begin{pmatrix} 0 & -1 \\ 0 & -2 \\ 1 & 0 \end{pmatrix}$ | $\begin{pmatrix} 0 & -2 \\ 0 & -1 \\ 1 & 0 \end{pmatrix}$ | $\begin{pmatrix} 0 & 1 \\ 0 & -1 \\ 1 & 0 \end{pmatrix}$ | $\begin{pmatrix} 0 & -1 \\ 0 & 1 \\ 1 & 0 \end{pmatrix}$ | $\begin{pmatrix} 0 & 2 \\ 0 & 1 \\ 1 & 0 \end{pmatrix}$ |
| $\Gamma_3$ | $\begin{pmatrix} 1 \\ 0 \\ 0 \end{pmatrix}$ | $\begin{pmatrix} 1 \\ 0 \\ 0 \end{pmatrix}$ | $\begin{pmatrix} 0 \\ 1 \\ 0 \end{pmatrix}$ | $\begin{pmatrix} -1 \\ -1 \\ 0 \end{pmatrix}$ | $\begin{pmatrix} -1 \\ -1 \\ 0 \end{pmatrix}$ | $\begin{pmatrix} 0 \\ 1 \\ 0 \end{pmatrix}$ |
| $\Gamma_4$ | $\begin{pmatrix} 0 & 1 \\ 0 & 2 \\ 1 & 0 \end{pmatrix}$ | $\begin{pmatrix} 0 & 1 \\ 0 & 2 \\ -1 & 0 \end{pmatrix}$ | $\begin{pmatrix} 0 & -2 \\ 0 & -1 \\ 1 & 0 \end{pmatrix}$ | $\begin{pmatrix} 0 & 1 \\ 0 & -1 \\ 1 & 0 \end{pmatrix}$ | $\begin{pmatrix} 0 & 1 \\ 0 & -1 \\ -1 & 0 \end{pmatrix}$ | $\begin{pmatrix} 0 & -2 \\ 0 & -1 \\ -1 & 0 \end{pmatrix}$ |
| $\Gamma_5$ | $\begin{pmatrix} 2 & 0 & 0 & 0 & 2 & 0 \\ 0 & 0 & 0 & 0 & 4 & 0 \\ 0 & 0 & 0 & 0 & 0 & 2 \end{pmatrix}$ | $\begin{pmatrix} -2 & 0 & 0 & 0 & -2 & 0 \\ 0 & 0 & 0 & 0 & -4 & 0 \\ 0 & -2 & 0 & 0 & 0 & 2 \end{pmatrix}$ | $\begin{pmatrix} 0 & 2 & 0 & 0 & 2 & 0 \\ -1 & 1 & 0 & 1 & 1 & 0 \\ 0 & 0 & -1 & 0 & 0 & -1 \end{pmatrix}$ | $\begin{pmatrix} 1 & 1 & 0 & 1 & -1 & 0 \\ 1 & -1 & 0 & 1 & 1 & 0 \\ 0 & 0 & 1 & 0 & 0 & -1 \end{pmatrix}$ | $\begin{pmatrix} -1 & -1 & 0 & -1 & 1 & 0 \\ -1 & 1 & 0 & -1 & -1 & 0 \\ 0 & 0 & 1 & 0 & 0 & -1 \end{pmatrix}$ | $\begin{pmatrix} 0 & -2 & 0 & 0 & -2 & 0 \\ 1 & -1 & 0 & -1 & -1 & 0 \\ 0 & 0 & -1 & 0 & 0 & -1 \end{pmatrix}$ |
| $\Gamma_6$ | $\begin{pmatrix} 2 & 0 & 0 & 0 & 0 & 2 \\ 4 & 0 & 0 & 0 & 0 & 0 \\ 0 & 2 & 0 & 0 & 0 & 0 \end{pmatrix}$ | $\begin{pmatrix} 2 & 0 & 0 & 0 & 0 & 2 \\ 4 & 0 & 0 & 0 & 0 & 0 \\ 0 & -2 & 0 & 0 & 0 & 0 \end{pmatrix}$ | $\begin{pmatrix} 2 & 0 & 0 & -2 & 0 & 0 \\ 1 & 0 & -1 & -1 & 0 & -1 \\ 0 & -1 & 0 & 0 & 1 & 0 \end{pmatrix}$ | $\begin{pmatrix} -1 & 0 & -1 & -1 & 0 & 1 \\ 1 & 0 & -1 & 1 & 0 & 1 \\ 0 & -1 & 0 & 0 & -1 & 0 \end{pmatrix}$ | $\begin{pmatrix} -1 & 0 & -1 & -1 & 0 & 1 \\ 1 & 0 & -1 & 1 & 0 & 1 \\ 0 & 1 & 0 & 0 & 1 & 0 \end{pmatrix}$ | $\begin{pmatrix} 2 & 0 & 0 & -2 & 0 & 0 \\ 1 & 0 & -1 & -1 & 0 & -1 \\ 0 & 1 & 0 & 0 & -1 & 0 \end{pmatrix}$ |
| $\Gamma_7$ | $\begin{pmatrix} 0 & 1 \\ 0 & 2 \\ 1 & 0 \end{pmatrix}$ | $\begin{pmatrix} 0 & -1 \\ 0 & -2 \\ 1 & 0 \end{pmatrix}$ | $\begin{pmatrix} 0 & -2 \\ 0 & -1 \\ 1 & 0 \end{pmatrix}$ | $\begin{pmatrix} 0 & 1 \\ 0 & -1 \\ 1 & 0 \end{pmatrix}$ | $\begin{pmatrix} 0 & -1 \\ 0 & 1 \\ 1 & 0 \end{pmatrix}$ | $\begin{pmatrix} 0 & 2 \\ 0 & 1 \\ 1 & 0 \end{pmatrix}$ |
| $\Gamma_8$ | $\begin{pmatrix} 1 \\ 0 \\ 0 \end{pmatrix}$ | $\begin{pmatrix} -1 \\ 0 \\ 0 \end{pmatrix}$ | $\begin{pmatrix} 0 \\ 1 \\ 0 \end{pmatrix}$ | $\begin{pmatrix} -1 \\ -1 \\ 0 \end{pmatrix}$ | $\begin{pmatrix} 1 \\ 1 \\ 0 \end{pmatrix}$ | $\begin{pmatrix} 0 \\ -1 \\ 0 \end{pmatrix}$ |
| $\Gamma_9$ | $\begin{pmatrix} 0 & 1 \\ 0 & 2 \\ 1 & 0 \end{pmatrix}$ | $\begin{pmatrix} 0 & 1 \\ 0 & 2 \\ -1 & 0 \end{pmatrix}$ | $\begin{pmatrix} 0 & -2 \\ 0 & -1 \\ 1 & 0 \end{pmatrix}$ | $\begin{pmatrix} 0 & 1 \\ 0 & -1 \\ 1 & 0 \end{pmatrix}$ | $\begin{pmatrix} 0 & 1 \\ 0 & -1 \\ -1 & 0 \end{pmatrix}$ | $\begin{pmatrix} 0 & -2 \\ 0 & -1 \\ -1 & 0 \end{pmatrix}$ |
| $\Gamma_{10}$ | $\begin{pmatrix} 1 \\ 0 \\ 0 \end{pmatrix}$ | $\begin{pmatrix} 1 \\ 0 \\ 0 \end{pmatrix}$ | $\begin{pmatrix} 0 \\ 1 \\ 0 \end{pmatrix}$ | $\begin{pmatrix} -1 \\ -1 \\ 0 \end{pmatrix}$ | $\begin{pmatrix} -1 \\ -1 \\ 0 \end{pmatrix}$ | $\begin{pmatrix} 0 \\ 1 \\ 0 \end{pmatrix}$ |
| $\Gamma_{11}$ | $\begin{pmatrix} 1 & -1 & 0 & 1 & -1 & 0 \\ 0 & 1 & 0 & 0 & -1 & 0 \\ 0 & 0 & 1 & 0 & 0 & -1 \end{pmatrix}$ | $\begin{pmatrix} -1 & 1 & 0 & -1 & 1 & 0 \\ 0 & -1 & 0 & 0 & 1 & 0 \\ 0 & 0 & 1 & 0 & 0 & -1 \end{pmatrix}$ | $\begin{pmatrix} 0 & 2 & 0 & 0 & 0 & 0 \\ 0 & 1 & 0 & -2 & 1 & 0 \\ 0 & 0 & -2 & 0 & 0 & 0 \end{pmatrix}$ | $\begin{pmatrix} 1 & -1 & 0 & -1 & 1 & 0 \\ 1 & -2 & 0 & -1 & 0 & 0 \\ 0 & 0 & 1 & 0 & 0 & 1 \end{pmatrix}$ | $\begin{pmatrix} -1 & 1 & 0 & 1 & -1 & 0 \\ -1 & 2 & 0 & 1 & 0 & 0 \\ 0 & 0 & 1 & 0 & 0 & 1 \end{pmatrix}$ | $\begin{pmatrix} 0 & -2 & 0 & 0 & 0 & 0 \\ 0 & -1 & 0 & 2 & -1 & 0 \\ 0 & 0 & -2 & 0 & 0 & 0 \end{pmatrix}$ |
| $\Gamma_{12}$ | $\begin{pmatrix} 2 & 0 & 0 & 0 & 2 & 0 \\ 0 & 0 & 0 & 0 & 4 & 0 \\ 0 & 0 & 0 & 0 & 0 & 2 \end{pmatrix}$ | $\begin{pmatrix} 2 & 0 & 0 & 0 & 2 & 0 \\ 0 & 0 & 0 & 0 & 4 & 0 \\ 0 & 0 & 0 & 0 & 0 & -2 \end{pmatrix}$ | $\begin{pmatrix} 0 & 2 & 0 & 0 & 2 & 0 \\ -1 & 1 & 0 & 1 & 1 & 0 \\ 0 & 0 & -1 & 0 & 0 & -1 \end{pmatrix}$ | $\begin{pmatrix} 1 & 1 & 0 & 1 & -1 & 0 \\ 1 & -1 & 0 & 1 & 1 & 0 \\ 0 & 0 & 1 & 0 & 0 & -1 \end{pmatrix}$ | $\begin{pmatrix} 1 & 1 & 0 & 1 & -1 & 0 \\ 1 & -1 & 0 & 1 & 1 & 0 \\ 0 & 0 & -1 & 0 & 0 & 1 \end{pmatrix}$ | $\begin{pmatrix} 0 & 2 & 0 & 0 & 2 & 0 \\ -1 & 1 & 0 & 1 & 1 & 0 \\ 0 & 0 & -1 & 0 & 0 & 1 \end{pmatrix}$ |

Table III(b): Basis vectors of the irreducible representation $\Gamma_n$ for the remaining six equivalent positions of Fe5 at the 12k site with fractional coordinate (x=0.168, y=2x, z=-0.1082).

| Magnetic Representation | Fe5 (12k) | | | | | |
|---|---|---|---|---|---|---|
| | $\begin{pmatrix} x-y \\ -y \\ -z \end{pmatrix}$ | $\begin{pmatrix} y \\ x \\ -z \end{pmatrix}$ | $\begin{pmatrix} -x \\ -x+y \\ -z \end{pmatrix}$ | $\begin{pmatrix} x \\ x-y \\ -z+3/2 \end{pmatrix}$ | $\begin{pmatrix} -x+y \\ y \\ -z+3/2 \end{pmatrix}$ | $\begin{pmatrix} -y \\ -x \\ -z+3/2 \end{pmatrix}$ |
| $\Gamma_1$ | $\begin{pmatrix} 1 \\ 0 \\ 0 \end{pmatrix}$ | $\begin{pmatrix} 0 \\ 1 \\ 0 \end{pmatrix}$ | $\begin{pmatrix} -1 \\ -1 \\ 0 \end{pmatrix}$ | $\begin{pmatrix} 1 \\ 1 \\ 0 \end{pmatrix}$ | $\begin{pmatrix} -1 \\ 0 \\ 0 \end{pmatrix}$ | $\begin{pmatrix} 0 \\ -1 \\ 0 \end{pmatrix}$ |
| $\Gamma_2$ | $\begin{pmatrix} 0 & 1 \\ 0 & 2 \\ 1 & 0 \end{pmatrix}$ | $\begin{pmatrix} 0 & -2 \\ 0 & -1 \\ 1 & 0 \end{pmatrix}$ | $\begin{pmatrix} 0 & 1 \\ 0 & -1 \\ 1 & 0 \end{pmatrix}$ | $\begin{pmatrix} 0 & -1 \\ 0 & 1 \\ 1 & 0 \end{pmatrix}$ | $\begin{pmatrix} 0 & -1 \\ 0 & -2 \\ 1 & 0 \end{pmatrix}$ | $\begin{pmatrix} 0 & 2 \\ 0 & 1 \\ 1 & 0 \end{pmatrix}$ |
| $\Gamma_3$ | $\begin{pmatrix} 1 \\ 0 \\ 0 \end{pmatrix}$ | $\begin{pmatrix} 0 \\ 1 \\ 0 \end{pmatrix}$ | $\begin{pmatrix} -1 \\ -1 \\ 0 \end{pmatrix}$ | $\begin{pmatrix} -1 \\ -1 \\ 0 \end{pmatrix}$ | $\begin{pmatrix} 1 \\ 0 \\ 0 \end{pmatrix}$ | $\begin{pmatrix} 0 \\ 1 \\ 0 \end{pmatrix}$ |
| $\Gamma_4$ | $\begin{pmatrix} 0 & -1 \\ 0 & -2 \\ -1 & 0 \end{pmatrix}$ | $\begin{pmatrix} 0 & 2 \\ 0 & 1 \\ -1 & 0 \end{pmatrix}$ | $\begin{pmatrix} 0 & -1 \\ 0 & 1 \\ -1 & 0 \end{pmatrix}$ | $\begin{pmatrix} 0 & 1 \\ 0 & -1 \\ -1 & 0 \end{pmatrix}$ | $\begin{pmatrix} 0 & 1 \\ 0 & 2 \\ -1 & 0 \end{pmatrix}$ | $\begin{pmatrix} 0 & -2 \\ 0 & -1 \\ -1 & 0 \end{pmatrix}$ |
| $\Gamma_5$ | $\begin{pmatrix} 2 & 0 & 0 & 0 & 2 & 0 \\ 0 & 0 & 0 & 0 & 4 & 0 \\ 0 & 0 & 0 & 0 & 0 & 2 \end{pmatrix}$ | $\begin{pmatrix} 0 & 2 & 0 & 0 & 2 & 0 \\ -1 & 1 & 0 & 1 & 1 & 0 \\ 0 & 0 & -1 & 0 & 0 & -1 \end{pmatrix}$ | $\begin{pmatrix} 1 & 1 & 0 & 1 & -1 & 0 \\ 1 & -1 & 0 & 1 & 1 & 0 \\ 0 & 0 & 1 & 0 & 0 & -1 \end{pmatrix}$ | $\begin{pmatrix} -1 & -1 & 0 & -1 & 1 & 0 \\ -1 & 1 & 0 & -1 & -1 & 0 \\ 0 & 0 & 1 & 0 & 0 & -1 \end{pmatrix}$ | $\begin{pmatrix} -2 & 0 & 0 & 0 & -2 & 2 \\ 0 & 0 & 0 & 0 & -4 & 0 \\ 0 & 0 & 0 & 0 & 0 & 0 \end{pmatrix}$ | $\begin{pmatrix} 0 & -2 & 0 & 0 & -2 & 0 \\ 1 & -1 & 0 & -1 & -1 & 0 \\ 0 & 0 & -1 & 0 & 0 & -1 \end{pmatrix}$ |
| $\Gamma_6$ | $\begin{pmatrix} 2 & 0 & 0 & 0 & 0 & 2 \\ 4 & 0 & 0 & 0 & 0 & 0 \\ 0 & 2 & 0 & 0 & 0 & 0 \end{pmatrix}$ | $\begin{pmatrix} 2 & 0 & 0 & -2 & 0 & 0 \\ 1 & 0 & -1 & -1 & 0 & -1 \\ 0 & -1 & 0 & 0 & 1 & 0 \end{pmatrix}$ | $\begin{pmatrix} -1 & 0 & -1 & -1 & 0 & 1 \\ 1 & 0 & -1 & 1 & 0 & 1 \\ 0 & -1 & 0 & 0 & -1 & 0 \end{pmatrix}$ | $\begin{pmatrix} -1 & 0 & -1 & -1 & 0 & 1 \\ 1 & 0 & -1 & 1 & 0 & 1 \\ 0 & 1 & 0 & 0 & 1 & 0 \end{pmatrix}$ | $\begin{pmatrix} 2 & 0 & 0 & 0 & 0 & 2 \\ 4 & 0 & 0 & 0 & 0 & 0 \\ 0 & -2 & 0 & 0 & 0 & 0 \end{pmatrix}$ | $\begin{pmatrix} 2 & 0 & 0 & -2 & 0 & 0 \\ 1 & 0 & -1 & -1 & 0 & -1 \\ 0 & 1 & 0 & 0 & -1 & 0 \end{pmatrix}$ |
| $\Gamma_7$ | $\begin{pmatrix} 0 & -1 \\ 0 & -2 \\ -1 & 0 \end{pmatrix}$ | $\begin{pmatrix} 0 & 2 \\ 0 & 1 \\ -1 & 0 \end{pmatrix}$ | $\begin{pmatrix} 0 & -1 \\ 0 & 1 \\ -1 & 0 \end{pmatrix}$ | $\begin{pmatrix} 0 & 1 \\ 0 & -1 \\ -1 & 0 \end{pmatrix}$ | $\begin{pmatrix} 0 & 1 \\ 0 & 2 \\ -1 & 0 \end{pmatrix}$ | $\begin{pmatrix} 0 & -2 \\ 0 & -1 \\ -1 & 0 \end{pmatrix}$ |
| $\Gamma_8$ | $\begin{pmatrix} -1 \\ 0 \\ 0 \end{pmatrix}$ | $\begin{pmatrix} 0 \\ -1 \\ 0 \end{pmatrix}$ | $\begin{pmatrix} 1 \\ 1 \\ 0 \end{pmatrix}$ | $\begin{pmatrix} -1 \\ -1 \\ 0 \end{pmatrix}$ | $\begin{pmatrix} 1 \\ 0 \\ 0 \end{pmatrix}$ | $\begin{pmatrix} 0 \\ 1 \\ 0 \end{pmatrix}$ |
| $\Gamma_9$ | $\begin{pmatrix} 0 & -1 \\ 0 & -2 \\ -1 & 0 \end{pmatrix}$ | $\begin{pmatrix} 0 & 2 \\ 0 & 1 \\ -1 & 0 \end{pmatrix}$ | $\begin{pmatrix} 0 & -1 \\ 0 & 1 \\ -1 & 0 \end{pmatrix}$ | $\begin{pmatrix} 0 & -1 \\ 0 & 1 \\ 1 & 0 \end{pmatrix}$ | $\begin{pmatrix} 0 & -1 \\ 0 & -2 \\ 1 & 0 \end{pmatrix}$ | $\begin{pmatrix} 0 & 2 \\ 0 & 1 \\ 1 & 0 \end{pmatrix}$ |
| $\Gamma_{10}$ | $\begin{pmatrix} -1 \\ 0 \\ 0 \end{pmatrix}$ | $\begin{pmatrix} 0 \\ -1 \\ 0 \end{pmatrix}$ | $\begin{pmatrix} 1 \\ 1 \\ 0 \end{pmatrix}$ | $\begin{pmatrix} 1 \\ 1 \\ 0 \end{pmatrix}$ | $\begin{pmatrix} -1 \\ 0 \\ 0 \end{pmatrix}$ | $\begin{pmatrix} 0 \\ -1 \\ 0 \end{pmatrix}$ |
| $\Gamma_{11}$ | $\begin{pmatrix} -1 & 1 & 0 & -1 & 1 & 0 \\ 0 & -1 & 0 & 0 & 1 & 0 \\ 0 & 0 & -1 & 0 & 0 & 1 \end{pmatrix}$ | $\begin{pmatrix} 0 & -2 & 0 & 0 & 0 & 0 \\ 0 & -1 & 0 & 2 & -1 & 0 \\ 0 & 0 & 2 & 0 & 0 & 0 \end{pmatrix}$ | $\begin{pmatrix} -1 & 1 & 0 & 1 & -1 & 0 \\ -1 & 2 & 0 & 1 & 0 & 0 \\ 0 & 0 & -1 & 0 & 0 & -1 \end{pmatrix}$ | $\begin{pmatrix} 1 & -1 & 0 & -1 & 1 & 0 \\ 1 & -2 & 0 & -1 & 0 & 0 \\ 0 & 0 & -1 & 0 & 0 & -1 \end{pmatrix}$ | $\begin{pmatrix} 1 & -1 & 0 & 1 & -1 & 0 \\ 0 & 1 & 0 & 0 & -1 & 0 \\ 0 & 0 & -1 & 0 & 0 & 1 \end{pmatrix}$ | $\begin{pmatrix} 0 & 2 & 0 & 0 & 0 & 0 \\ 0 & 1 & 0 & -2 & 1 & 0 \\ 0 & 0 & 2 & 0 & 0 & 0 \end{pmatrix}$ |
| $\Gamma_{12}$ | $\begin{pmatrix} -2 & 0 & 0 & 0 & -2 & 0 \\ 0 & 0 & 0 & 0 & -4 & 0 \\ 0 & 0 & 0 & 0 & 0 & -2 \end{pmatrix}$ | $\begin{pmatrix} 0 & -2 & 0 & 0 & -2 & 0 \\ 1 & -1 & 0 & -1 & -1 & 0 \\ 0 & 0 & 1 & 0 & 0 & 1 \end{pmatrix}$ | $\begin{pmatrix} -1 & -1 & 0 & -1 & 1 & 0 \\ -1 & 1 & 0 & -1 & -1 & 0 \\ 0 & 0 & -1 & 0 & 0 & 1 \end{pmatrix}$ | $\begin{pmatrix} -1 & -1 & 0 & -1 & 1 & 0 \\ -1 & 1 & 0 & -1 & -1 & 0 \\ 0 & 0 & 1 & 0 & 0 & -1 \end{pmatrix}$ | $\begin{pmatrix} -2 & 0 & 0 & 0 & -2 & 0 \\ 0 & 0 & 0 & 0 & -4 & 0 \\ 0 & 0 & 0 & 0 & 0 & 2 \end{pmatrix}$ | $\begin{pmatrix} 0 & -2 & 0 & 0 & -2 & 0 \\ 1 & -1 & 0 & -1 & -1 & 0 \\ 0 & 0 & -1 & 0 & 0 & -1 \end{pmatrix}$ |

Table IV: Bond lengths of Fe in kagome bilayer configuration linked via pyrochlore slabs.

| | |
|---|---|
| $Fe_{12k}$-$Fe_{12k}$ | 2.9156(0) Å |
| $Fe_{12k}$-$Fe_{12k}$ | 2.9686(7) Å |
| $Fe_{12k}$-$Fe_{2a}$ | 3.037592) Å |
| $Fe_{12k}$-$Fe_{4f1}$ | 3.4783(7) Å |
| $Fe_{12k}$-$Fe_{4f1}$ | 3.4786(8) Å |
| $Fe_{12k}$-$Fe_{4f2}$ | 3.4941(3) Å |
| $Fe_{12k}$-$Fe_{4f2}$ | 3.4943(8) Å |
| $Fe_{12k}$-$Fe_{2b}$ | 3.6972(1) Å |